\documentclass{article}

\usepackage[utf8]{inputenc}

\usepackage{authblk}

\usepackage{natbib}

\usepackage{amsmath,amsfonts,amssymb,amsthm,epsfig,epstopdf,url,array}
\usepackage{amsmath,amsthm}
\theoremstyle{plain}
\newtheorem{theorem}{Theorem}[section]
 
\newtheorem{lemma}[theorem]{Lemma}
\newtheorem{corollary}[theorem]{Corollary}
\newtheorem{remark}{Remark}[section]

\theoremstyle{definition}
\newtheorem{exmp}{Example}[section]

\newcommand*{\Cdot}{\raisebox{-0.25ex}{\scalebox{1.2}{$\cdot$}}}

\usepackage[title]{appendix}

\usepackage{tikz,lipsum,lmodern}
\usepackage[most]{tcolorbox}
\usepackage{smartdiagram}
\usepackage{amsfonts}
\usepackage{amssymb}
\usepackage{amsthm}
\usepackage{amsmath}
\usepackage{amsopn}
\usepackage{bm}
\usepackage{amsmath}
\usepackage{graphicx}
\usepackage{float}
\usepackage[utf8]{inputenc}
\usepackage{multirow}
\usepackage{rotating}
\usepackage{graphicx}
\usepackage{subcaption}
\usepackage{caption}
\usepackage{diagbox}
\usepackage{placeins}
\usepackage{xcolor}
\usepackage{soul}

\numberwithin{equation}{section}

\begin{document}

\title{Optimal Design in Repeated Testing for Count Data}

\author[a]{Parisa Parsamaram\footnote{Corresponding author, email: \texttt{parisaparsamaramf@gmail.com>}}} 
\author[b]{Heinz Holling}
\author[a]{Rainer Schwabe}
\affil[a]{Otto-von-Guericke University, Institute for Mathematical Stochastics, Universit\"atsplatz 2, 39106 Magdeburg, Germany}
\affil[b]{University of M\"unster, Institute for Psychology, Fliednerstra{\ss}e 21, 48149 M\"unster, Germany}

\date{}                     
\maketitle
\section*{Abstract}
\begin{abstract}
	In this paper, we develop optimal designs for growth curve models with count data based on the Rasch Poisson-Gamma counts (RPGCM) model. 
	This model is often used in educational and psychological testing when test results yield count data. 
	In the RPGCM, the test scores are determined by respondents ability and item difficulty. 
	Locally $D$-optimal designs are derived for maximum quasi-likelihood estimation to efficiently estimate the mean abilities of the respondents over time. 
	Using the log link, both unstructured, linear and nonlinear growth curves of log mean abilities are taken into account. 
	Finally, the sensitivity of the derived optimal designs due to an imprecise choice of parameter values is analyzed using $D$-efficiency.
	\vspace{3mm}

	\noindent
	Keywords: 
	D-optimal design, Rasch Poisson-Gamma counts model, growth curve model, quasi-likelihood estimation
\end{abstract}

\section{Introduction}

The analysis of retest effects in cognitive abilities has long been an important topic in education, psychology, and neuroscience. 
Retest effects, also known as test or practice effects, are defined as the change in test scores as a result of repeating the same or alternative cognitive ability tests under comparable conditions. 
As a recent meta-analysis by \cite{scharfen2018meta-analysis} shows, retests of the most important cognitive abilities lead to significant score increases that are usually nonlinear over the first test administrations before reaching a plateau. 
In addition, the mean increase in score gains differs considerably between subjects. 
Such different growth curves need to be taken into account to appropriately interpret test results, especially when testing experience varies between test takers.

Optimal design can be very useful in reducing the number of test items presented to test takers in repeating testing. 
Thus, \cite{freise2023repeated} developed optimal designs for estimating growth curves of cognitive ability tests with metric scores.  
However, many tests of cognitive abilities, particularly those that measure divergent thinking, memory, or processing speed, result in count data. 
In this case, linear models with normally distributed random effects and errors, as used by \cite{freise2023repeated}, are no longer appropriate. 

A frequently used model for cognitive tests with counting data is the Rasch Poisson-Gamma counts model \citep{grasshoff2020rasch}, an extension of the first item response model developed by the Danish mathematician \cite{Rasch60}. 
According to the RPGCM, the item scores consisting of the sum of correctly processed stimuli (e.g. correctly memorized words) are assumed to follow a Poisson distribution with intensity (mean) $\theta \sigma$, i.\,e.\ the product of the ability parameter $\theta$ of the respondent and the item parameter $\sigma$ which is defined as the easiness of the test item. 
A gamma distribution is assumed for the individual ability of the respondents, which leads to the Rasch Poisson-Gamma counts model \citep {Grasshoffetal10}. 
So far, this model has been specified for one point in time and will be extended below for repeated testing.

For the longitudinal Rasch Poisson-Gamma counts model, we will derive $D$-optimal designs to efficiently estimate the mean ability parameter by specifying the optimal number of items presented to the respondents at each time point. 
These plans may contradict the common practice in repeated testing where the same number of items are applied at each test time point, but may be more efficient. 

Since the likelihood of the longitudinal model involves complicated functions
defined by integrals which cannot be solved explicitly, maximum quasi-likelihood estimation based on the marginal first and second order moments of the response variable is used to estimate the parameters \citep[see][]{wedderburn1974quasi}. 
The performance of the design is then measured in terms of the quasi-information which plays the same role for maximum quasi-likelihood estimation as the Fisher information does for classical maximum likelihood \citep[see][Chapter~9]{McCullagh1989generalized2ed}.
The concept of quasi-information has been employed in the context of optimal design in the past e.\,g.\ by~\cite{niaparast2013optimal}, \cite{shen2016quasi}, and \cite{shi2023penalizedquasi} for random coefficient models in the presence of normally distributed random effects.
In contrast to these papers, the structure of the random effects is less restrictive in the present work and the random effects have a conjugate distribution as discussed in \cite{molenberghs2011conjugate}.

The structure of the present paper is as follows: 
We specify the longitudinal Rasch Poisson-Gamma counts model in Section~\ref{sec:model} and the maximum quasi-likelihood estimation of the parameters in the subsequent Section~\ref{sec:estimation}. 
Locally $D$-optimal designs for various kinds of growth curves are derived in Section~\ref{sec:design} and several examples are presented in Section~\ref{sec:examples}. 
In Section~\ref{sec:discussion}, a final discussion concludes this article. 
Technical proofs are deferred to the Appendix.

\section{Model specification}
\label{sec:model}
Similar to \cite{freise2023repeated}, we consider the situation of repeated
testing where a total number of $n$ items is presented 
to each of $N$ test persons over a period of time in order to investigate 
the mean learning curve in retesting.
More specifically, tests may be conducted at $J$ subsequent instances, 
$t_1 < \ldots < t_J$, and, at time point $t_j$, each test person $i$ ($i = 1, \ldots, N$)
receives the same number $n_j$ of items, $\sum_{j = 1}^{J} n_j = n$.
We denote by $y_{ijk}$ the response of test person $i$ to the $k$th item presented 
at time point $t_j$.

In \cite{freise2023repeated}, 
the outcome $y_{ijk}$ was assumed to be measured on a metric scale.
In contrast to that, we start from count data as outcomes. 
The distribution of the corresponding response variable $Y_{ijk}$
may then be described by the Rasch Poisson counts model 
\citep[see][and literature cited therein]{grasshoff2020rasch}.
There, on the single response level, the response $Y_{ijk}$
is assumed to follow a Poisson distribution with intensity (mean) 
$\theta_{ij} \sigma_{jk}$,
where $\theta_{ij}$ is the ability of person $i$ at test instance $j$
and $\sigma_{jk}$ is the easiness of the $k$th item presented at
that instance.
Within a test person, the responses $Y_{ijk}$
are assumed to be independent. 
In the test situation considered here, interest is in the ability parameters 
while the easiness of the items are supposed to be known. 

We will be interested in the development of the ability over time
in the sense of a common mean learning curve $\theta(t)$
for all test persons.
Then it is reasonable to build up a hierarchical model
where the subject (and instance) specific deviations
in the ability can be described by random effects
when the test persons come from a homogeneous population
such that $\theta_{ij}$ is the outcome of a latent variable $\Theta_{ij}$.
This will be assumed here throughout.

In contrast to the approach of normally distributed random effects
often used in life sciences, 
we suppose multiplicative gamma distributed random effects 
as in \cite{grasshoff2020rasch},
see also \cite{molenberghs2011conjugate}.
More specifically, we assume a multiplicative variation in
the ability such that $\theta_{ij} = \lambda_{ij} \theta(t_j)$
for test person $i$ at the $j$th instance,
and the factor $\lambda_{ij}$ is the outcome 
of a gamma distributed latent variable $\Lambda_{ij}$
such that $\Theta_{ij} = \Lambda_{ij} \theta(t_j)$.

The marginal mean response $\mathrm{E}(Y_{ijk})$ can be obtained as 
\begin{equation}
	\label{eq:marginal-mean}
	\mathrm{E}(Y_{ijk}) 
	= \mathrm{E}\left(\mathrm{E}(Y_{ijk} \, | \, \Lambda_{ij})\right) 
	= \mathrm{E}(\Lambda_{ij}) \theta(t_j) \sigma_{jk} \, .
\end{equation}
To achieve that the growth curve $\theta(t)$
represents the marginal mean ability by 
$\mathrm{E}(\Theta_{ij}) = \theta(t_j)$,
the variational factor $\Lambda_{ij}$ has to be normalized 
with standardized expectation, $\mathrm{E}(\Lambda_{ij}) = 1$.
\vspace{2mm}

On the within test person level, we allow the random effects 
$\Lambda_{i1}, \ldots , \Lambda_{iJ}$ to be correlated,
but require that they are exchangeable over time
which means, in particular, that they are identically distributed and equally correlated.
This condition corresponds to the concept of compound symmetry
for repeated measurements in linear mixed models
with multivariate normal random effects.
The exchangeability can be accomplished by the introduction of 
a permanent test person effect $\Gamma_{i0}$
and time specific person effects $\Gamma_{ij}$
for each time point $t_j$, $j  = 1, \ldots , J$, which, together,
constitute additively the random effects, 
$\Lambda_{ij} = \Gamma_{i0} + \Gamma_{ij}$,
see \cite{mathai1991multivariate}, under the following assumptions:
The time specific person effects $\Gamma_{i1}, \ldots , \Gamma_{iJ}$ 
are independent and identically gamma distributed with 
shape parameter $\alpha \geq 0$ and scale parameter $\tau > 0$.
The permanent test person effect $\Gamma_{i0}$
is gamma distributed with 
shape parameter $\alpha_0 \geq 0$ and the same scale parameter $\tau$
as the time specific person effects.
Moreover, the permanent test person effect $\Gamma_{i0}$ 
and the time specific person effects 
$\Gamma_{i1}, \ldots , \Gamma_{iJ}$, are independent. 
Then, the random effects $\Lambda_{ij} = \Gamma_{i0} + \Gamma_{ij}$ 
are gamma distributed with 
shape parameter $\alpha_0 + \alpha$ and scale parameter $\tau$,
and the vector $\bm{\Lambda}_{i} = (\Lambda_{i1}, \ldots ,\Lambda_{iJ})^{\top}$
of random effects follows a multivariate gamma distribution
with correlated components \citep{mathai1991multivariate}. 
The components $\Lambda_{ij}$ of $\bm{\Lambda}_i$
have mean $\mathrm{E}(\Lambda_{ij}) = (\alpha_0 + \alpha)\tau$,
variance $\mathrm{Var}(\Lambda_{ij}) = (\alpha_0 + \alpha)\tau^2$, 
covariance $\mathrm{cov}(\Lambda_{ij},\Lambda_{ij^{\prime}}) 
= \mathrm{Var}(\Gamma_{i0}) = \alpha_0\tau^2$, 
and, hence, correlation 
$\rho = \mathrm{corr}(\Lambda_{ij},\Lambda_{ij^{\prime}}) 
= \alpha_0/(\alpha_0 + \alpha)$, 
$j, j^{\prime} = 1, \ldots, J$, $j \neq j^{\prime}$.
In this modeling, independent time effects ($\rho = 0$) and 
a pure permanent test person effect ($\rho = 1$) 
are included as limiting cases, 
when $\alpha_0 = 0$, i.\,e.\ $\Lambda_{ij} = \Gamma_{ij}$,
and $\alpha = 0$, i.\,e.\ $\Lambda_{ij} = \Gamma_{i0}$,
respectively.
For this, note that a shape parameter equal to zero corresponds 
to a degenerate distribution with all mass concentrated at zero.

To attain the standardization $\mathrm{E}(\Lambda_{ij}) = 1$
for the mean of the random effects $\Lambda_{ij}$, 
the shape parameters $\alpha$ and $\alpha_0$ are chosen to satisfy 
$\alpha_0 + \alpha = 1/\tau$.
Under this condition, the random effects $\Lambda_{ij}$ have
variance $\mathrm{Var}(\Lambda_{ij}) = \tau$
and covariance 
$\mathrm{cov}(\Lambda_{ij},\Lambda_{ij^{\prime}}) = \rho\tau$, 
where $\rho = \alpha_0\tau$ is the intraclass correlation. 
The variance $\tau$ of the random effects
will play a similar role as the variance ratio does in the  
linear longitudinal model in~\cite{freise2023repeated}.
In total, the vector $\bm{\Lambda}_{i}$
of random effects is multivariate gamma
with mean vector $\mathrm{E}(\bm{\Lambda}_i) = \mathbf{1}_J$
and covariance matrix 
\begin{equation}
	\label{eq:cov-lambda}
	\mathrm{Cov}(\bm{\Lambda}_i) 
	= \tau \left((1 - \rho) \mathbf{I}_J + \rho \mathbf{1}_J \mathbf{1}_J^\top\right) \, ,
\end{equation}
where $\mathbf{I}_m$ and $\mathbf{1}_m$ denote the
$m \times m$ identity matrix and the $m$-dimensional vector
with all entries equal to one, respectively.

For further use, we denote by
$\mathbf{Y}_{ij} = (Y_{ij1}, \ldots, Y_{ijn_{j}})^{\top}$
the $n_j$-dimensional vector of responses at  time $t_j$ and by
$\mathbf{Y}_{i} = (\mathbf{Y}_{i1}^{\top}, \ldots, \mathbf{Y}_{iJ}^{\top})^{\top}$
the $n$-dimensional vector of all responses of test person $i$.
Note that $\mathbf{Y}_{ij}$ may be void when 
no items were presented at time $t_j$ ($n_j = 0$).
Moreover, to be more precise, we assume that
all responses $Y_{ijk}$ are independent
given the full vector $\bm{\Lambda}_i$
of random effects.
\vspace{2mm}

On the population level, we suppose that the random effects 
$\bm{\Lambda}_i$ and, hence, the test results $\mathbf{Y}_i$ 
are independent between test persons
and that the mean ability is suitably parameterized.
As common in generalized linear (mixed) models
for count data, we use a log link for the mean ability
such that $\theta(t) = \exp(\eta(t, \bm{\beta}))$,
where the functional relationship $\eta(t, \bm{\beta})$
is known up to a $p$-dimensional parameter vector 
$\bm{\beta} = (\beta_1, \ldots, \beta_p)^\top$, $p \leq J$.
Then, for the individual ability, we have
$\theta_{ij} = \exp(\eta(t_{ij}, \bm{\beta}) + \log(\lambda_{ij}))$
and the standardization $\mathrm{E}(\Lambda_{ij}) = 1$
of the random effect $\Lambda_{ij}$ may be interpreted 
as an analogue to the requirement of zero expectation 
for normally distributed random effects in linear models.
\vspace{2mm}

\textbf{Unstructured (anova type) mean ability. }
If no particular form is assumed 
how the mean ability $\theta(t)$ evolves over time, 
the functional relationship may be parameterized
by \begin{equation}
	\label{eq:model-unstructured}
	\eta(t_j, \bm{\beta}) = \beta_j \, , \qquad j = 1, \ldots, p = J \, ,
\end{equation}
such that $\theta(t_j) = \log(\beta_j)$.
\vspace{2mm}

\textbf{(Generalized) Linear model type log mean ability. }
Similar to the situation in a generalized linear model,
a more structured functional relationship may be obtained
when the growth curve $\theta(t) = \exp(\eta(t, \bm{\beta}))$ 
is described by a linear component 
\begin{equation}
	\label{eq:model-linear}
	\eta(t, \bm{\beta}) = \mathbf{f}(t)^\top \bm{\beta} \, ,
\end{equation}
where $\mathbf{f}(t) = (f_1(t), \ldots, f_p(t))^\top$ is a 
$p$-dimensional vector of (known) regression functions
$f_1(t), \ldots, f_p(t)$.

For example, a straight line (ordinary linear regression type) 
relationship can be given by 
\begin{equation}
	\label{eq:model-straight-line}
	\eta(t, \bm{\beta}) = \beta_1 + \beta_2 t 
\end{equation}
such that $f_1(t) \equiv 1$ is constant, $f_2(t) = t$ is the
identity, and the dimension is $p = 2$. 
In this case, the parameters $\beta_1$ and $\beta_2$ denote
the intercept and slope parameter, respectively.
\vspace{2mm}

The unstructured case~\eqref{eq:model-unstructured} is included 
in the linear model type case~\eqref{eq:model-linear}
when dummy coding is used as in an anova setting, 
i.\,e.\ $f_j(t) = 1$ if $t = t_j$, and $f_j(t) = 0$ otherwise.
\vspace{2mm}

\textbf{Nonlinear mean ability. }
To model certain properties occurring in practice like saturation,
a nonlinear functional relationship $\eta(t, \bm{\beta})$ 
will occur.
For asymptotic properties, we suppose that the functional relationship
$\eta(t, \bm{\beta})$ is smooth in the sense that it is differentiable
with respect to the parameters $\bm{\beta}$.
We denote by
\begin{equation}
	\label{eq:model-nonlinear}
	\mathbf{f}_{\bm{\beta}}(t) 
		= \frac{\partial}{\partial \bm{\beta}} \eta(t, \bm{\beta})
\end{equation}
the gradient vector of $\eta(t, \bm{\beta})$ with respect to $\bm{\beta}$.
Then $\mathbf{f}_{\bm{\beta}}(t)^\top \bm{\beta}$ is a first order 
approximation of $\eta(t, \bm{\beta})$.

As an example, we consider the 
three parameter growth curve
 \begin{equation}
 	\label{eq:model-growth}
	\eta(t, \bm{\beta}) = \beta_{1} - \beta_{2} \exp (- \beta_{3} t) \, .
\end{equation}
When the range parameter $\beta_{2}$
and the scale parameter $\beta_{3}$
are positive, 
the log mean ability increases in time 
and approaches the
saturation level $\beta_{1}$ for large $t$. 
At initial time $t = 0$, the log mean ability is equal to
$\beta_1 - \beta_2$ such that the range parameter $\beta_{2}$ 
is the maximal amount of gain in log mean ability.
Finally, the (inverse) scale parameter $\beta_{3}$ indicates the speed of 
growth in log mean ability.
In particular, the log mean ability  
approaches the saturation level faster 
for larger values of the scale parameter $\beta_{3}$.
Hence, this exponential saturation model
is a reasonable alternative for the
commonly used 4PL (E-max) model
when only observations in the upper tail 
will be available.

For the exponential saturation model, the gradient vector is
\begin{equation}
	\label{eq:gradient-growth}
	\mathbf{f}_{\bm{\beta}}(t) 
	= (1, - \exp(- \beta_{3} t), \beta_{2} t \exp(- \beta_{3} t))^{\top} 
\end{equation}
and varies with the range and scale parameter
$\beta_2$ and $\beta_3$.
\vspace{2mm}

In contrast to formula~\eqref{eq:gradient-growth}
for the (nonlinear) exponential saturation model,
the gradient  
$\frac{\partial}{\partial \bm{\beta}} \eta(t, \bm{\beta}) 
= \mathbf{f}(t)$
does not depend on the parameter vector $\bm{\beta}$
in the linear model type case~\eqref{eq:model-linear}.

\section{Estimation and information}
\label{sec:estimation}
Interest is in estimating the population parameters 
$\bm{\beta}$ which determine the log mean ability curve
$\eta(t, \bm{\beta})$ and, hence, the mean ability curve
$\theta(t)$ .
For simplification, we assume 
for the random effects $\bm{\Lambda}_i$
that the scale parameter
$\tau$ and the intraclass correlation $\rho$  are known.
Then the marginal density of the responses $\mathbf{Y}_i$
and, hence, the likelihood may be expressed explicitly 
\citep[see][Chapter 5, Appendix]{parsamaram2022thesis}.
However, this expression involves complicated functions
defined by infinite series and cannot directly be used for 
determining the maximum likelihood estimate of $\bm{\beta}$.
To circumvent this problem, we make use of the concept of
maximum quasi-likelihood estimation suggested by 
\cite{wedderburn1974quasi}. 
For this kind of estimation, only the first and second order
marginal moments of the response vector $\mathbf{Y}_i$ 
are required rather than the full marginal density of $\mathbf{Y}_i$. 

Following~\cite{wedderburn1974quasi}, 
we define the maximum quasi-likelihood estimator 
$\hat{\bm{\beta}}_{\mathrm{MQL}}$ of $\bm{\beta}$
as a solution of the generalized estimating equation
\begin{equation}
	\label{eq:mql-estimate}
	\sum_{i=1}^{N} \mathbf{D}_i^\top \mathbf{V}_i^{-1}
		(\mathbf{Y}_i - \bm{\mu}_i)
	= \mathbf{0} \, ,
\end{equation}
where 
$\bm{\mu}_i = \mathrm{E}(\mathbf{Y}_i)$
and
$\mathbf{V}_i = \mathrm{Cov}(\mathbf{Y}_i)$
are the marginal mean and marginal covariance matrix 
of the responses $\mathbf{Y}_i$ of test person $i$,
and $\mathbf{D}_i =
\frac{\partial}{\partial \bm{\beta}^\top} \bm{\mu}_i$
is the matrix of partial derivatives
of the marginal mean $\bm{\mu}_i$
with respect to the components of $\bm{\beta}$. 
Note that the marginal covariance matrix $\mathbf{V}_i$,
the marginal mean $\bm{\mu}_i$ and, hence, the matrix
$\mathbf{D}_i$ are functions of the parameter vector  $\bm{\beta}$.
To compute $\hat{\bm{\beta}}_{\mathrm{MQL}}$ 
a modified Fisher scoring method is suggested \cite[see][]{liang1986longitudinal}. 
The maximum quasi-likelihood estimator $\hat{\bm{\beta}}_{\mathrm{MQL}}$ 
may be interpreted as a simplified version of a
generalized least squares estimator
$\hat{\bm{\beta}}_{\mathrm{GLS}} 
= \arg\min
	\sum_{i=1}^{N} (\mathbf{Y}_i - \bm{\mu}_i)^\top 
	\mathbf{V}_i^{-1} (\mathbf{Y}_i - \bm{\mu}_i)$,
where the dependence of the marginal covariance matrix 
$\mathbf{V}_i$ on $\bm{\beta}$ is neglected
when taking the derivatives of
$(\mathbf{Y}_i - \bm{\mu}_i)^\top \mathbf{V}_i^{-1} (\mathbf{Y}_i - \bm{\mu}_i)$
with respect to the components of $\bm{\beta}$.

Similar to the Fisher information, the quasi-information 
\begin{equation}
	\label{eq:quasi-information}
	\mathbf{M}_{\mathrm{Q}}
	= \sum_{i=1}^{N} \mathbf{D}_{i}^{\top} \mathbf{V}_{i}^{-1} \mathbf{D}_{i}
\end{equation}
is defined as the covariance matrix of the quasi-score function
given by the left hand side 
$\sum_{i=1}^{N} \mathbf{D}_i^\top \mathbf{V}_i^{-1} (\mathbf{Y}_i - \bm{\mu}_i)$
of equation~\eqref{eq:mql-estimate}
\citep[see][]{McCullagh1989generalized2ed}.
Under certain regularity conditions,
the asymptotic covariance of the maximum quasi-likelihood estimator 
is proportional to the inverse of the quasi-information matrix.
In particular, the quasi-information matrix has to be nonsingular.
The quasi-information plays the same role 
for the quasi-likelihood estimator
as the Fisher information does
for the common maximum-likelihood estimator,
and the quality of performance of the
maximum quasi-likelihood estimator can be measured by 
the quasi-information matrix.

In the present setting, the test conditions are the same
for all test persons.
Hence, $\bm{\mu}_i = \bm{\mu}$, $\mathbf{V}_i =\mathbf{V}$,
$\mathbf{D}_i = \mathbf{D}$, and
the generalized estimating equations~\eqref{eq:mql-estimate}
for the maximum quasi-likelihood estimator 
$\hat{\bm{\beta}}_{\mathrm{MQL}}$ reduce to
$\mathbf{D}^\top \mathbf{V}^{-1} (\bar{\mathbf{Y}} - \bm{\mu}) 
= \mathbf{0}$, where
$\bar{\mathbf{Y}} = \frac{1}{N} \sum_{i=1}^{N} \mathbf{Y}_i$
is the average of the individual responses,
and the quasi-information~\eqref{eq:quasi-information} simplifies to 
\begin{equation}\label{eq:quasi-onformation-equal}
	\mathbf{M}_{\mathrm{Q}}
	= N \, \mathbf{D}^{\top} \mathbf{V}^{-1} \mathbf{D} \, .
\end{equation}

Denote by $\mu_{jk} = \theta(t_{j})\sigma_{jk}$
the marginal mean $\mathrm{E}(Y_{ijk})$ 
of the response $Y_{ijk}$ 
to the $k$th item at time $t_j$.
Further, denote now by 
$\bm{\mu}_j = (\mu_{j1}, \ldots, \mu_{jn_{j}})^{\top}$ 
the marginal mean $\mathrm{E}(\mathbf{Y}_{ij})$
of the responses $\mathbf{Y}_{ij}$ at time $t_j$.
Then
$\bm{\mu} = (\bm{\mu}_{1}^{\top}, \ldots, \bm{\mu}_{J}^{\top})^{\top}$
is the common marginal mean of all responses $\mathbf{Y}_i$
for all test persons.
As a consequence, the common
$n \times p$ matrix $\mathbf{D}$ of partial derivatives is given by
\begin{equation}
	\label{eq:matrix-d-varying-easiness}
	\mathbf{D} 
	= (\mathbf{D}_{1}^{\top}, \ldots , \mathbf{D}_{J}^{\top})^{\top},
\end{equation}
where 
$\mathbf{D}_{j} = (\mathbf{d}_{j1}, \ldots , \mathbf{d}_{jn_{j}})^{\top}$
and $\mathbf{d}_{jk} = \frac{\partial \mu_{jk}}{\partial \bm{\beta}}$
is the gradient vector of the mean response $\mu_{jk}$ 
for the $k$th item at time $t_j$ with respect to $\bm{\beta}$.
By~\eqref{eq:model-nonlinear},
the gradient can be written as
$\mathbf{d}_{jk} = \mu_{jk} \mathbf{f}_{\bm{\beta}}(t_j)$.

For the second order moments, we make use of the 
conditional mean and variance 
$\mathrm{Var}(Y_{ijk} | \bm{\Lambda}_i) 
= \mathrm{E}(Y_{ijk} | \bm{\Lambda}_i) = \Lambda_{ij} \mu_{jk}$
and the conditional independence of all $Y_{ijk}$ given $\bm{\Lambda}_i$.
Then we get
\begin{equation}
	\label{eq:var-y_ijk}
	\mathrm{Var}(Y_{ijk}) 
	=
	\mathrm{Var}\left(\mathrm{E}(Y_{ijk} | \bm{\Lambda}_i)\right)
	+ \mathrm{E}\left(\mathrm{Var}(Y_{ijk} | \bm{\Lambda}_i)\right)
	= \tau \mu_{jk}^2 + \mu_{jk} 
\end{equation}
for the marginal variance of the response $Y_{ijk}$,
\begin{equation}
	\label{eq:cov-y_ijk-y_ijl}
	\mathrm{cov}(Y_{ijk}, Y_{ijk^{\prime}})
	=
	\mathrm{cov}\left(\mathrm{E}(Y_{ijk} | \bm{\Lambda}_i), 
	\mathrm{E}(Y_{ijk^{\prime}} | \bm{\Lambda}_i)\right)
	= \tau \mu_{jk} \mu_{jk^{\prime}} \, ,\ k\neq k^{\prime} ,
\end{equation}
for the marginal covariance of responses within time $t_j$, and
\begin{equation}
	\label{eq:cov-y_ijk-y_ihl}
	\mathrm{cov}(Y_{ijk},Y_{ij^{\prime}k^{\prime}})
	=
	\mathrm{cov}\left(\mathrm{E}(Y_{ijk} | \bm{\Lambda}_i), 
	\mathrm{E}(Y_{ij^{\prime}k^{\prime}} | \bm{\Lambda}_i)\right)
	= \rho \tau \mu_{jk} \mu_{j^{\prime}k^{\prime}} 
\end{equation}
for the marginal covariance of responses at different times 
$t_j$ and $t_{j^\prime}$.
The marginal covariance matrix $\mathbf{V} = \mathrm{Cov}(\mathbf{Y}_i)$
for all responses $\mathbf{Y}_i$ of a test person can, hence,
be written as
\begin{equation}
	\label{eq:covariance-y_i-varying-easiness}
	\mathbf{V} 
	= \mathrm{diag}\left(\mu_{jk}\right) 
	+ (1 - \rho) \tau \, \mathrm{diag}\left(\bm{\mu}_{j} \bm{\mu}_{j}^\top\right) 
	+ \rho \tau \bm{\mu}\bm{\mu}^{\top},
\end{equation}
where $\mathrm{diag}(\cdot)$ denotes a (block) diagonal matrix 
with the corresponding arguments on the diagonal.

To get more insight into the structure of the individual 
quasi-information matrix $\mathbf{D}^{\top} \mathbf{V}^{-1} \mathbf{D}$
associated with a single test person, 
we introduce the essential regression matrix 
\begin{equation}
	\label{eq:regression-matrix-essential}
	\mathbf{F} = (\mathbf{f}_{\bm{\beta}}(t_1), \ldots, \mathbf{f}_{\bm{\beta}}(t_J))^{\top}
\end{equation} 
of size $J \times p$
which consists of the regression vectors 
$\mathbf{f}_{\bm{\beta}}(t_j)$ for all $J$ time points,
irrespectively of the numbers $n_j$ of items.
Further, we define the quantity  
\begin{equation}
	\label{eq:psi}
	u_{j} = \frac{\mu_{j \Cdot}}{1 + (1 - \rho) \tau \mu_{j \Cdot}}
\end{equation}
which will cover the impact of the total mean response 
$\mu_{j \Cdot} = \sum_{k=1}^{n_j} \mu_{jk}$ 
at time $t_j$ on the quasi-information, 
the corresponding vector
$\mathbf{u} = (u_1, \ldots, u_J)^\top$ and
the diagonal matrix $\mathbf{U} = \mathrm{diag}(u_j)$.

Then, the individual quasi-information matrix
can be written as follows.
\begin{theorem}
	\label{thm:quasi-information}
	\begin{equation}
		\label{eq:quasi-information-representation}
		\mathbf{D}^{\top} \mathbf{V}^{-1} \mathbf{D}
		=
		\mathbf{F}^\top \mathbf{U} \mathbf{F} 
		- {\textstyle{\frac{\rho \tau}{1 + \rho \tau \sum_{j=1}^{J} u_{j}}}} 
		\mathbf{F}^\top \mathbf{u} \mathbf{u}^\top \mathbf{F} \, .
	\end{equation}
\end{theorem}

\begin{remark}
	The representation in Theorem~\ref{thm:quasi-information} 
	is valid even 
	when some of the numbers $n_j$ of items and, hence, 
	some $u_j$ are equal to zero.
\end{remark}

\begin{remark}
	Both the individual quasi-information matrix
	$\mathbf{D}^{\top} \mathbf{V}^{-1} \mathbf{D}$
	and the leading term $\mathbf{F}^\top \mathbf{U} \mathbf{F}$
	on the right hand side 
	of equation~\eqref{eq:quasi-information-representation}
	are nonsingular if and only if the design matrix $\mathbf{D}$
	is of full column rank $p$.
\end{remark}

\section{Design}
\label{sec:design}
The performance of the maximum quasi-likelihood estimator
depends on the choice of the experimental conditions.
The aim of determining a good design 
is to maximize quasi-information 
in order to minimize the asymptotic covariance 
of the quasi-likelihood estimator.
We first note that the quasi-information
gets larger when easier items are used.
\begin{lemma}
	\label{lem:max-easiness}
	Let $n_{1}, \ldots , n_{J}$ be fixed.
	Then the quasi-information
	is increasing in the easiness of the items
	in the Loewner sense of nonnegative definiteness.
\end{lemma}
Hence, the quasi-information is maximized
if all items are adjusted to maximal easiness,
$\sigma_{jk} = \sigma$, say, for all $j$ and $k$.
This will be assumed henceforward.

Then, in the present situation 
of fixed time points $t_1, \ldots , t_J$
of testing and a fixed total number $n$ of items,
the design of the experiment is completely described
by the vector $(n_1, \ldots , n_J)$ of numbers of items presented
at time $t_1, \ldots , t_J$, $n_j \geq 0$, $\sum_{j=1}^J n_j = n$,
and the performance of the design
can be measured in
terms of the quasi-information matrix
$\mathbf{M}_{\mathrm{Q}}(n_1, \ldots, n_J)
= N \mathbf{D}^{\top} \mathbf{V}^{-1} \mathbf{D}$.
As this maximization cannot be done with respect to
the Loewner ordering of matrices in the sense
of nonnegative domination,
we will adopt here the $D$-criterion 
$\log(\det(\mathbf{M}_{\mathrm{Q}}(n_1, \ldots, n_J)))$
of maximizing the determinant of the quasi-information matrix,
\citep[see e.\,g.][]{silvey1980optimal}.
The $D$-criterion is the most popular criterion in the literature
and may be interpreted as minimization of the volume
of the asymptotic confidence ellipsoid based on the
maximum quasi-likelihood estimator.

Moreover, it has to be noted that the quasi-information
is also a function of $\bm{\beta}$
as common in nonlinear and generalized linear models.
Hence, the $D_{\mathrm{Q}}$-optimal design which maximizes the $D$-criterion
may depend on the value of the parameter vector $\bm{\beta}$.
We, thus, consider local $D_{\mathrm{Q}}$-optimality with respect to
some prespecified initial guess for the parameter value $\bm{\beta}$
\citep[see][]{atkinson2007optimum}.

To find a (locally) $D_{\mathrm{Q}}$-optimal design 
in the class of exact designs $(n_1, \ldots, n_J)$
under the constraints $n_j \geq 0$,  $j = 1, \ldots, J$,
and $\sum_{j=1}^J n_j = n$
is a discrete optimization problem
which is hard to handle.
To facilitate optimization,
we follow the concept of approximate designs
introduced by \cite{kiefer1959optimal}
and allow a relaxation of the numbers $n_j$
to be continuous.

More precisely, we define an (approximate) design
$\xi$ as a vector $\xi = (w_1, \ldots, w_J)$ of weights
satisfying $w_j \geq 0$,  $j = 1, \ldots, J$,
and $\sum_{j=1}^J w_j = 1$.
For an exact design $(n_1, \ldots, n_J)$,
the weights $w_j = n_j / n$
of the corresponding approximate design
$\xi = (n_1 / n, \ldots, n_J / n)$
may be interpreted as the proportions of items presented
at time points $t_1, \ldots, t_J$, respectively.
However, for a general approximate design,
we permit the weights $w_j$ to be continuous.
We may then use tools from (convex) analysis
to solve the optimization problem
to obtain (locally) $D_{\mathrm{Q}}$-optimal designs 
$\xi^* = (w_1^*, \ldots , w_J^*)$.

To embed the discrete optimization problem 
into the continuous context, we make use of
the representation~\eqref{eq:quasi-information-representation}
of the quasi-information matrix 
in Theorem~\ref{thm:quasi-information}.
In this representation, the quasi-information
depends on the design only through the quantities $u_j$
defined by~equation\eqref{eq:psi}.

When all items have the same easiness $\sigma$,
then the mean response $\mu_{jk}$ 
is constant for all items
presented at the same instance $j$,
$\mu_{jk} = \mu_{j} = \theta(t_{j}) \sigma$
say, and 
$u_{j} = \frac{n_{j} \mu_{j}}{1 + (1 - \rho) \tau n_{j} \mu_{j}}$.
This expression can readily be extended to
approximate designs.
For this, we introduce the function 
\begin{equation}
	\label{eq:psi-func}
	u_{j}(w) = \frac{\mu_{j} w}{1 + (1 - \rho) n \tau \mu_{j} w} \, .
\end{equation}
For any design  $\xi = (w_1, \ldots, w_J)$, let 
$\mathbf{U}(\xi) = \mathrm{diag}(u_j(w_j))$
be the corresponding diagonal matrix 
of weighting factors $u_j(w_j)$ and 
\begin{equation}
	\label{eq:matrix-m}
	\mathbf{M}(\xi) = \mathbf{F}^\top \mathbf{U}(\xi) \mathbf{F} 
\end{equation}
corresponding to the leading term on the right hand side
of~\eqref{eq:quasi-information-representation}.
With this notation, we define the
standardized (per item) quasi-information
\begin{equation}
	\label{eq:quasi-information-xi}
	\mathbf{M}_{\mathrm{Q}}(\xi) 
	= \mathbf{M}(\xi) 
		- {\textstyle{\frac{\rho n \tau}{1 + \rho n \tau \mathbf{1}_J^\top \mathbf{U}(\xi) \mathbf{1}_J}}} 
			\mathbf{F}^\top \mathbf{U}(\xi) \mathbf{1}_J \mathbf{1}_J^\top \mathbf{U}(\xi) \mathbf{F}
\end{equation}
for the design $\xi$.
Then, in view of the 
representation~\eqref{eq:quasi-information-representation} 
in Theorem~\ref{thm:quasi-information}, 
it can be seen that 
$\mathbf{M}_{\mathrm{Q}}(\xi) 
	= \frac{1}{Nn} \mathbf{M}_{\mathrm{Q}}(n_1, \ldots, n_J)$
for the approximate design $\xi = (n_1 / n, \ldots, n_J / n)$ 
corresponding to an exact design $(n_1, \ldots, n_J)$
such that formula~\eqref{eq:quasi-information-xi} provides 
a proper generalization of the quasi-information.
An approximate design 
$\xi^* = (w_1^*, \ldots, w_J^*)$
is then said to be $D_{\mathrm{Q}}$-optimal if it
maximizes the $D$-criterion
$\Phi_{D, \mathrm{Q}}(\xi) 
	= \log(\det(\mathbf{M}_{\mathrm{Q}}(\xi)))$
in the class of all approximate designs.
If the optimal weights $w_j^*$ result
in integer solutions $n_j^* = n w_j^*$,
then the corresponding exact design
$(n_1^*, \ldots, n_J^*)$ will be $D_{\mathrm{Q}}$-optimal itself.
Otherwise, optimal (or, at least, nearly optimal)
exact designs can be obtained 
by suitable rounding of the numbers $n w_j^*$
to integer values when the total number $n$ of items
per test person is sufficiently large.

\begin{remark}
	\label{rem:quasi-information-fixed}
	In the case of a fixed effect Poisson model with independent
	observations $Y_{ijk}$, $\Lambda_{ij} \equiv 1$,
	the weighting functions reduce to 
	$u_j(w) = \mu_j w$
	and the quasi-information to
	\begin{equation}
		\label{eq:quasi-information-fixed}
		\mathbf{M}_{\mathrm{Q}}(\xi) 
			= \mathbf{M}(\xi) 
			= \mathbf{F}^\top \mathrm{diag}(\mu_j w_j) \mathbf{F} \, .
	\end{equation}
	can be obtained by formally letting
	$\tau = 0$ in \eqref{eq:quasi-information-xi}.
	
	In the fixed effects model,
	both the maximum quasi-likelihood estimator
	and the quasi-information matrix
	coincide with the maximum-likelihood estimator
	and the Fisher information matrix,
	respectively.
		
	Moreover, in the fixed effects model, 
	the per item quasi-information
	$\mathbf{M}_{\mathrm{Q}}(\xi)$
	given in~\eqref{eq:quasi-information-fixed} 
	does not depend on the total number $n$ of items
	and, hence, the per test person
	quasi-information $n \mathbf{M}_{\mathrm{Q}}(\xi)$
	increases linearly in $n$.
	In contrast to that, in the mixed effect model, 
	the per item quasi-information 
	decreases in $n$ while the per test person
	quasi-information 
	becomes stable because of the test person 
	specific random effects ($\tau > 0$).
\end{remark}

\begin{remark}
	\label{rem:quasi-information-rho0}
	In the special case $\rho = 0$ 
	of independent time effects,
	the quasi-information matrix 
	\[
		\mathbf{M}_{\mathrm{Q}}(\xi) = \mathbf{M}(\xi) \, .
	\]
	simplifies to the leading term 
	in the definition~\eqref{eq:quasi-information-xi}
	with $u_{j}(w) = \mu_{j} w / (1 + n \tau \mu_{j} w)$.
	Then the $D$-criterion reduces to
	$\Phi_{D, \mathrm{Q}}(\xi) = \log(\det(\mathbf{M}(\xi)))$.
	%
\end{remark}

In general, the $D$-criterion can be calculated as
\begin{eqnarray}
	\lefteqn{
		\Phi_{D, \mathrm{Q}}(\xi) 
		= 
		\log(\det(\mathbf{M}(\xi)))
			- \log\left(1 + \rho n \tau \mathbf{1}_J^\top \mathbf{U}(\xi) \mathbf{1}_J\right)
	}
	\nonumber
	\\
	&& \mbox{}
	+ \log\left(1 
		+ \rho n \tau \mathbf{1}_J^\top (\mathbf{U}(\xi)  
			-  \mathbf{U}(\xi) \mathbf{F} \mathbf{M}(\xi)^{-1}
				\mathbf{F}^\top \mathbf{U}(\xi)) \mathbf{1}_J\right) \, .
	\label{eq:d-crit-general}
\end{eqnarray}
by application of the common matrix determinant 
lemma~\ref{lem:matrix-determinant-lemma} 
to the definition given in equation~\eqref{eq:quasi-information-xi}. 
This formula is difficult to handle 
for optimization purposes.
Therefore, we will look for some simplification
which can be achieved when the last term 
vanishes in~\eqref{eq:d-crit-general}.

In particular, if there is correlation over time ($\rho > 0$), 
then there is a permanent test person effect $\Gamma_{i0}$
which may be considered 
as some kind of random intercept.
Therefore, it seems to be reasonable that the 
functional relationship $\eta(t, \bm{\beta})$ 
itself contains a fixed effect intercept.
More precisely, we assume that 
the constant function $\equiv 1$ is in the span of the regression functions
$\mathbf{f}_{\bm{\beta}}$, i.\,e.\ there exists a vector $\mathbf{c}$
such that
$\mathbf{F} \mathbf{c} = \mathbf{1}_{J}$. 
Then we may obtain a more useful representation 
of the quasi-information matrix
$\mathbf{M}_{\mathrm{Q}}(\xi)$
when the quasi-information or,
equivalently, the matrix $\mathbf{M}(\xi)$
is nonsingular.
\begin{theorem}
	\label{thm:quasi-information-intercept}
	Let $\mathbf{F} \mathbf{c} = \mathbf{1}_J$
	and let $\mathbf{M}(\xi)$ be nonsingular.
	Then, for the design $\xi$, the quasi-information matrix is
	\begin{equation}
		\label{eq:quasi-information-intercept}
		\mathbf{M}_{\mathrm{Q}}(\xi)
			= \left(\mathbf{M}(\xi)^{-1} 
				+ \rho n \tau \mathbf{c} \mathbf{c}^{\top}\right)^{-1} .
	\end{equation}
\end{theorem}

The condition $\mathbf{F} \mathbf{c} = \mathbf{1}_{J}$ 
is met in many situations, in particular,
when there is an explicit intercept,
$\beta_1$ say, as in model~\eqref{eq:model-nonlinear}
such that $f_1(t) \equiv 1$ and $\mathbf{c} = \mathbf{e}_1$ 
is the unit vector with first entry equal to one.

\begin{remark}
	\label{rem:intercept-implicit}
	The condition of an (implicit) intercept 
	is automatically satisfied  
	when the model is saturated, i.\,e.\ $J = p$
	and $\mathbf{F}$ has full rank.
	In particular,
	in the unstructured case~\eqref{eq:model-unstructured},
	this holds with $\mathbf{F} = \mathbf{I}_J$
	and $\mathbf{c} = \mathbf{1}_J$.
	There the representation~\eqref{eq:quasi-information-intercept}
	of the quasi-information matrix becomes

	\begin{equation}
		\label{eq:quasi-information-unstructured}
		\mathbf{M}_{\mathrm{Q}}(\xi)
			= \left(
					\mathrm{diag}(1 / (\mu_j w_j))
						+ n \tau 
							\left(
								(1 - \rho) \mathbf{I}_J 
								+ \rho \mathbf{1}_J \mathbf{1}_J^{\top}
							\right)
				\right)^{-1} .
	\end{equation}
\end{remark}

\begin{remark}
	\label{rem:quasi-information-rho1}
	In the special case $\rho = 1$ of a pure permanent test 
	person effect,
	the weighting functions $u_j(w) = \mu_j w$
	do not depend on the scale parameter $\tau$
	and the total number $n$ of items.
	However, the quasi-information~\eqref{eq:quasi-information-xi} 
	still does by the leading factor 
	$n \tau / (1 + n \tau \sum_{j = 1}^{J} \mu_j w_j)$
	of its second term.
	Further, in the presence of an intercept, 
	$\mathbf{F} \mathbf{c} = \mathbf{1}_J$,
	the representation~\eqref{eq:quasi-information-intercept} 
	of the quasi-information
	\begin{equation}
		\label{eq:quasi-information-rho1}
		\mathbf{M}_{\mathrm{Q}}(\xi)
			= \left((\mathbf{F}^\top \mathrm{diag}(\mu_j w_j) \mathbf{F})^{-1}
				+ n \tau \mathbf{c} \mathbf{c}^{\top}\right)^{-1}
	\end{equation}
	coincides with the form of the Fisher information matrix
	obtained by~\citet[Lemma~3.3]{schmidt2020block}.
	The leading term $\mathbf{M}(\xi)$
	in~\eqref{eq:quasi-information-rho1}
	is equal to the quasi-information
	in the fixed effect model~\eqref{eq:quasi-information-fixed}.
\end{remark}

Concerning the $D$-criterion,
the last term of the
representation~\eqref{eq:d-crit-general} 
vanishes 
under the condition of an intercept.

\begin{corollary}
	\label{cor:d-criterion-intercept}
	Let $\mathbf{F} \mathbf{c} = \mathbf{1}_{J}$.
	Then the $D$-criterion is of the form
	\begin{equation}
		\label{eq:d-criterion-intercept}
		\Phi_{D, \mathrm{Q}}(\xi) 
		= \log\left(\det(\mathbf{M}(\xi))\right) 
			- \log\left(1 + \rho n \tau \mathbf{1}_{J}^{\top} \mathbf{U}(\xi) \mathbf{1}_{J}\right) \, .
	\end{equation}
\end{corollary}

As already mentioned, the $D$-criterion 
depends on the vector $\bm{\beta}$ 
of location parameters through $\mathbf{U}$
via the mean responses $\mu_j$
and potentially through the linearized
regression $\mathbf{f}_{\bm{\beta}}$
in the essential regression matrix 
$\mathbf{F}$
when the log mean ability curve $\eta(t)$
is nonlinear.
Additionally, the $D$-criterion also 
depends on the number $n$ of observations
per subject, the scale parameter $\tau$
and the correlation $\rho$.
The dependence on $n$ and $\tau$
is only through the size related scale 
$a = n \tau$ per subject
which plays a similar role as the standardized
variance ratio in the linear longitudinal model
in~\cite{freise2023repeated}.
Thus, also the $D_{\mathrm{Q}}$-optimal
design may depend on all these quantities.

\begin{remark}
	\label{rem:dcrit-Jequalp}
	When the model is saturated ($J = p$
	and $\mathbf{F}$ has full rank), 
	the optimal design depends on 
	the location parameter $\bm{\beta}$ 
	only through the mean responses $\mu_j$,
	irrespectively of the model described
	by the regression matrix $\mathbf{F}$.
\end{remark}

For $\rho = 0$ 
or $\mathbf{F} \mathbf{c} = \mathbf{1}_{J}$,
the $D$-criterion
$\Phi_{D, \mathrm{Q}}(\xi) 
	= \log(\det(\mathbf{M}_{\mathrm{Q}}(\xi)))$
is concave (see Lemma~\ref{lem:concave}).
We can thus apply convex analysis
to obtain characterizations for
optimal designs.

For this, denote by
$\mathbf{W}(\xi) = \mathrm{diag}(w_j)$ 
the diagonal matrix of the weights $w_j$ of $\xi$,
by $u_j^{\prime}(w) = \frac{\mu_j}{(1 + (1 - \rho) n \tau \mu_j w)^2}$
the derivative of $u_j(w)$,
and by $\mathbf{U}^{\prime}(\xi) = \mathrm{diag}(u_j^{\prime}(w_j))$
the corresponding diagonal matrix of the derivatives $u_j^{\prime}$ 
evaluated at the weights $w_j$ of $\xi$.

\begin{theorem}[Equivalence Theorem]
	\label{thm:equivalence}
	{\em (a)}
	Let $\rho = 0$.
	Then $\xi^{*}$ is locally $D_{\mathrm{Q}}$-optimal 
	at $\bm{\beta}$ if and only if 
	\begin{equation}
		u_j^{\prime}(w_j^*)
			\mathbf{f}_{\bm{\beta}}(t_j)^\top \mathbf{M}(\xi^*)^{-1} \mathbf{f}_{\bm{\beta}}(t_j)
		\leq 
		\mathrm{tr}\left(\mathbf{M}(\xi^*)^{-1} \mathbf{F}^\top
		\mathbf{W}(\xi^*) \mathbf{U}^{\prime}(\xi^*)
		\mathbf{F}\right)
		\label{eq:equivalence-rho0}
	\end{equation}
	for all $j = 1, \ldots, J$.
	\\
	{\em (b)}
	Let $\mathbf{F}\mathbf{c} = \mathbf{1}_{J}$.
	Then $\xi^{*}$ is locally $D_{\mathrm{Q}}$-optimal 
	at $\bm{\beta}$ if and only if 
	\begin{eqnarray}
		\lefteqn{u_j^{\prime}(w_j^*)
			\left(\mathbf{f}_{\bm{\beta}}(t_j)^\top \mathbf{M}(\xi^*)^{-1} \mathbf{f}_{\bm{\beta}}(t_j)
				- \frac{\rho n \tau}{1 + \rho n \tau \mathbf{1}_J^\top \mathbf{U}(\xi^*) \mathbf{1}_J}\right)}
		\label{eq:equivalence-intercept}
		\\
		\nonumber
		& \leq &
		\mathrm{tr}\left(\mathbf{M}(\xi^*)^{-1} \mathbf{F}^\top
				\mathbf{W}(\xi^*) \mathbf{U}^{\prime}(\xi^*) \mathbf{F}\right)
			- \frac{\rho n \tau \mathbf{1}_J^\top \mathbf{W}(\xi^*) \mathbf{U}^{\prime}(\xi^*) \mathbf{1}_J}{1 
				+ \rho n \tau \mathbf{1}_J^\top \mathbf{U}(\xi^*) \mathbf{1}_J} 
	\end{eqnarray}
	for all $j = 1, \ldots, J$.
\end{theorem}

\begin{remark}
	For $D_{\mathrm{Q}}$-optimal designs $\xi^*$, 
	equality holds in \eqref{eq:equivalence-rho0}
	and \eqref{eq:equivalence-intercept}, 
	respectively, for all $j$ 
	with positive weight $w_{j}^*$ at $t_j$. 

	In particular, when the model is saturated ($J = p$), 
	all weights $w_j^*$ have to be positive
	and, hence, equality holds  
	in~\eqref{eq:equivalence-rho0}
	and in~\eqref{eq:equivalence-intercept}, 
	respectively, for all $j$.
\end{remark}

To measure the sensitivity of the locally
$D_{\mathrm{Q}}$-optimal design,
we make use of the common concept of efficiency.
For any design $\xi$, the
$D_{\mathrm{Q}}$-efficiency 
is defined by
\begin{equation}
	\label{eq:effciency}
	\mathrm{eff}_{D, \mathrm{Q}}(\xi, \bm{\beta})
		= \left(\frac{\det(\mathbf{M}_{\mathrm{Q}}(\xi, \bm{\beta}))}
				{\det(\mathbf{M}_{\mathrm{Q}}(\xi_{\bm{\beta}}^{*}, \bm{\beta}))}
			\right)^{1/p},
\end{equation}
where, $\xi_{\bm{\beta}}^*$ denotes the locally 
$D_{\mathrm{Q}}$-optimal design at $\bm{\beta}$. 
For clarity of notation, we have added here the dependence 
of the quasi-information matrix 
on the parameter value $\bm{\beta}$ 
as an additional argument in  
$\mathbf{M}_{\mathrm{Q}}(\xi, \bm{\beta})$.
The efficiency of a design $\xi$ may be interpreted 
as the proportion of subjects needed 
when the optimal design 
$\xi_{\bm{\beta}}^*$ would be used
to achieve the same accuracy as under the design $\xi$.
Equivalently, the deficiency 
$\frac{1}{\mathrm{eff}_{D, \mathrm{Q}}(\xi)} - 1$
is the proportion of additional units needed 
under the design $\xi$
to achieve the same accuracy as under 
the optimal design $\xi_{\bm{\beta}}^*$.
For example, if the efficiency of $\xi$ is $0.5$, 
we would have to take twice as many subjects 
as for the optimal design $(\xi_{\bm{\beta}}^{*})$.

\section{Examples}
\label{sec:examples}

For illustrative purposes, we start with 
the unstructured 
case~\eqref{eq:model-unstructured}.

\begin{exmp}[unstructured, equal means]
	\label{exmp:unstructured-noeffect}
	In the unstructured 
	case~\eqref{eq:model-unstructured},
	there is no learning effect
	if all components of  $\bm{\beta}$ are equal
	($\beta_1 = \ldots = \beta_J$)
	such that the mean ability $\theta(t)$
	is constant over time.
	Then the time points $t_j$ are exchangeable,
	and the $D$-criterion is invariant with respect
	to permutations of the time points.
	Hence, in view of the strict concavity of the $D$-criterion,
	the uniform design $\bar{\xi}$
	which assigns equal weights $w_j = 1/J$ 
	to each time point is the (unique) locally $D_{\mathrm{Q}}$-optimal design.
	\qed
\end{exmp}

By Example~\ref{exmp:unstructured-noeffect},
the commonly used strategy
to present the same number of items
at each time point
is optimal when there is no learning  
(or fatigue) effect.
This is no longer true 
when the mean ability varies over time
as can be seen in the next example.

\begin{exmp}[unstructured, $J = 2$]
	\label{exmp:unstructured-J2}
	In the unstructured 
	case~\eqref{eq:model-unstructured}
	with $J = 2$ time points,
	the determinant of the inverse quasi-information 
	matrix~\eqref{eq:quasi-information-unstructured} is
	\begin{equation}
		\label{eq:det-unstructured-J2}
		\det(\mathbf{M}_{\mathrm{Q}}(\xi)^{-1})
			= \left(1 / (\mu_1 w_1) + n \tau\right)
				 \left(1 / (\mu_2 w_2) + n \tau\right)
				 - (\rho n \tau)^2 .
	\end{equation}
	Maximization of the $D$-criterion is equivalent 
	to minimization of the first term on the right hand side 
	of equation~\eqref{eq:det-unstructured-J2}.
	Hence, the $D_{\mathrm{Q}}$-optimal design
	does not depend on the correlation $\rho$.
	It depends on the other model parameters
	only through the quantities $a_j = a \mu_j$
	which describe the mean response $\mu_j$ 
	per item at instance $j$
	standardized by the size related scale 
	$a = n \tau$ per subject.
	By solving the optimization problem, 
	the optimal weights are obtained as
	$w_1^* 
		= \sqrt{a_2 + 1} / (\sqrt{a_1 + 1} + \sqrt{a_2 + 1})$
	and $w_2^* = 1 - w_1^*$ for both instances 
	of testing, respectively.
	These weights coincide with the optimal solutions
	obtained by~\citet[Theorem~5.3, p~35]{schmidt2020diss}
	for the Poisson model 
	with gamma distributed block effects ($\rho = 1$).
	
	In Figure~\ref{fig:unstructured-j2-weight-beta},
	the optimal weight $w_1^*$ is plotted 
	in dependence on the gain $\beta_2 - \beta_1$
	in log mean ability from the first to the second 
	instance of testing.
	\begin{figure}
		\centering
		\includegraphics[width=0.55\textwidth]{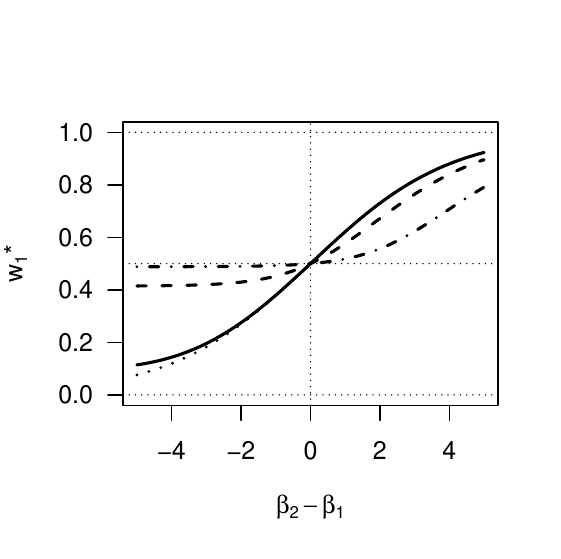}
		\caption{Optimal weight $w_1^*$
			in dependence on $\beta_2 - \beta_1$
			in the unstructured (anova) model ($p = J = 2$)
			for various values of $a_1 = n \tau \sigma \exp(\beta_1)$:
			$a_1 = 100$ (solid), $a_1 = 1$ (dashed), $a_1 = 0.1$ (dashed-dotted), 
			and $a_1 \to \infty$ (dotted)}
		\label{fig:unstructured-j2-weight-beta}
	\end{figure}
	The vertical dotted line at
	$\beta_2 - \beta_1 = 0$ indicates the situation of
	equal mean ability at both instances 
	which corresponds to the situation of
	Example~\ref{exmp:unstructured-noeffect}.
	There the uniform design $\bar{\xi}$
	with equal weights $w_{1}^* = w_{2}^* = 1/2$ 
	is optimal for any value of $a_1$.
	The uniform design $\bar{\xi}$ is further optimal
	when there are no random effects 
	($\tau = 0$ and, hence, $a_1 = 0$)
	as indicated by the horizontal dotted line at
	$w_1^* = 0.5$.
	
	For fixed gain $\beta_2 - \beta_1$
	(and fixed $n$ and $\tau$),
	the uniform design $\bar{\xi}$ also occurs as a
	limiting case when $\beta_1$ tends to $- \infty$
	and, hence, $a_1$ and $a_2$ tend to $0$.
	On the other hand, when $\beta_1$ and,
	hence, $a_1$ and $a_2$ tend to 
	$\infty$, the limiting curve is the logistic 
	function $1 / (1 + \exp( - (\beta_2 - \beta_1)))$.
	
	For fixed value of $a_1 > 0$,
	the optimal weight $w_1^*$ is increasing 
	in $\beta_2$.
	In the case of a positive gain ($\beta_2 > \beta_1$),
	the optimal weight $w_1^*$ is larger than $1/2$ 
	and $w_1^*$ tends to $1$ for $\beta_2 \to \infty$
	while in the case of fatigue ($\beta_2 < \beta_1$),
	$w_1^*$ is smaller than $1/2$ 
	and $w_1^*$ tends to 
	$1 / (\sqrt{a_1 + 1} + 1) > 0$ 
	for $\beta_2 \to - \infty$.
	In particular, the optimal design assigns more
	weight $w_j^*$ to that instance $j$
	for which the mean ability is smaller. 
			
	In the present setting,
	for any design $\xi = (w_1,  w_2)$,
	the efficiency~\eqref{eq:effciency} is
	\[
		\mathrm{eff}_{D, \mathrm{Q}}(\xi, \bm{\beta})
		=
		\left(
			\frac{\left(
						\sqrt{(a_1 + 1) (a_2 + 1)} + 1
					\right)^2 
						- \rho^2 a_1 a_2}
					{(a_1 + 1 / w_1) (a_2 + 1 / w_2) 
						- \rho^2 a_1 a_2}
		\right)^{1/2} .
	\]
	In contrast to the optimal weights, the efficiency
	$\mathrm{eff}_{D, \mathrm{Q}}(\xi, \bm{\beta})$
	depends also on the correlation $\rho$.
	This is illustrated in 
	Figure~\ref{fig:unstructured-j2-efficiency-beta}
	for the efficiency 
	$\mathrm{eff}_{D, \mathrm{Q}}(\bar{\xi}, \bm{\beta})$
	of the uniform design $\bar{\xi}$ 
	which assigns equal weight $w_j = 1/2$ 
	to each instance of testing. 
	\begin{figure}
		\centering
		\includegraphics[width=0.55\textwidth]{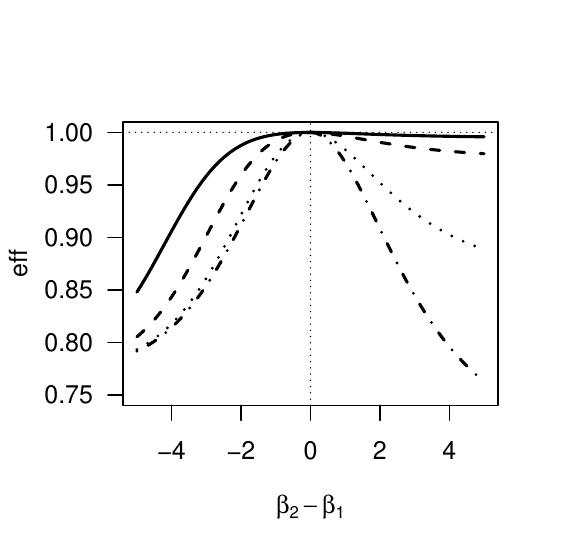}
		\caption{Efficiency of the uniform design 
			($w_1 = w_2 = 1/2$)
			in dependence on $\beta_2 - \beta_1$
			in the unstructured (anova) model ($p = J = 2$)
			for $a_1 = 100$ and 
			$\rho = 0$ (solid), $\rho = 0.9$ (dashed), $\rho = 0.99$ (dotted), 
			and $\rho = 1$ (dashed-dotted)}
		\label{fig:unstructured-j2-efficiency-beta}
	\end{figure}
	There the efficiency is plotted in dependence 
	on the gain $\beta_2 - \beta_1$ in log mean ability
	for $a_1 = 100$ and various values of $\rho$.
	The vertical dotted line at
	$\beta_2 - \beta_1 = 0$ indicates the situation of
	equal mean ability at both instances 
	for which the uniform design is optimal.
	This is reflected by the fact 
	that the efficiency is equal to $1$
	as indicated by the horizontal dotted line
	at $\mathrm{eff} = 1$.
	For fixed value of $\rho$,
	the efficiency decreases to
	$(((1 - \rho^2) a_1 + 1) / ((1 - \rho^2) a_1 + 2))^{1/2}$
	when the gain in log mean ability gets large
	($\beta_2 \to \infty$). 
	On the other hand,
	the efficiency decreases to
	$((1 / 2 + \sqrt{a_1 + 1} / (a_1 + 2)))^{1/2}$
	when there is a strong loss in log mean ability 
	($\beta_2 \to - \infty$). 
	Further, the efficiency decreases in 
	the correlation $\rho$
	as exemplified in 
	Figure~\ref{fig:unstructured-j2-efficiency-beta}.
	Thus efficiency is highest
	in the case of independent time effects ($\rho = 0$), 
	and it is lowest
	in the case of a pure permanent test person effect 
	($\rho = 1$).
	In the latter case, the efficiency decreases to
	$1/\sqrt{2} \approx 0.7071$ for large $\beta_2$
	while, for other values of the correlation ($\rho > 0$),
	the limiting efficiency may be substantially larger.
	For $\beta_2 \to - \infty$, the limiting efficiency
	does not depend on $\rho$ and is, 
	for the particular value $a_1 = 100$ in
	Figure~\ref{fig:unstructured-j2-efficiency-beta},
	equal to $0.7736$.
	For other values of $a_1$, the efficiency
	shows a similar behavior as depicted in
	Figure~\ref{fig:unstructured-j2-efficiency-beta}.
	In any case, the efficiency remains larger than 
	$70\,\%$ for all parameter combinations.
	\qed
\end{exmp}

For saturated models ($J = p$), the locally 
$D_{\mathrm{Q}}$-optimal design does not depend 
on the particular parameterization of the model 
but only on the mean responses $\mu_j$ 
--- apart from the total number $n$ of items, 
the scale parameter $\tau$ and the correlation $\rho$ 
of the random effects. 
Hence the results of the previous examples
apply also to other model parameterizations.
This property will be made use of in the 
subsequent Examples~\ref{exmp:straight-line-J3} 
and \ref{exmp:exponential-saturation}
when $J = p$.

However, these examples are primarily devoted to
nonsaturated model situations 
in which the number $p$ of parameters is less 
than the number $J$ of testing instances.
We start with the total mean only model 
in which it is known a priori 
that the mean ability does not change over time.
This is the most simple nonsaturated model 
which fulfills the condition 
$\mathbf{F}\mathbf{c} = \mathbf{1}_{J}$ 
of containing an intercept.
There the mean ability is the same as in 
Example~\ref{exmp:unstructured-noeffect},
but the prior knowledge of constant mean ability 
can be employed
which will lead to a different design problem.

\begin{exmp}[total mean only]
	\label{exmp:intercept-only}
	If the mean ability $\theta(t)$ 
	is known to be constant over time,
	then the model consists only of an intercept
	($p = 1$, $f(t) \equiv 1$, $\theta = \exp(\beta)$).
	As in 
	Example~\ref{exmp:unstructured-noeffect},
	the time points are exchangeable,
	and the quasi-information is invariant 
	with respect to permutations of the time points.
	Hence, by concavity,
	the uniform design $\bar{\xi}$ is optimal ($w_j^* = 1/J$)
	for every value of $n$, $\theta$, $\tau$, and $\rho$.
	When $\rho < 1$, the quasi-information 
	is strictly concave and $\bar{\xi}$ 
	is the unique optimal design.
	If $\rho = 1$, 
	observations are indistinguishable and
	the quasi-information $M_{\mathrm{Q}}(\xi) = \mu$
	is constant in $\xi$
	such that all designs are optimal.
	\qed
\end{exmp}

Next, we consider the case
of a linear trend
in log mean ability over time.

\begin{exmp}[straight line log mean ability]
	\label{exmp:straight-line-J3}
	The situation~\eqref{eq:model-straight-line}
	 of a linear development 
	$\eta(t, \bm{\beta}) = \beta_1 + \beta_2 t$
	of the log mean ability
	is characterized by 
	the regression function 
	$\mathbf{f}(t) = (1, t)^\top$ 
	and $p = 2$ parameters.
	For $J = 2$ time points,
	the model is saturated,
	and locally $D_{\mathrm{Q}}$-optimal designs can be obtained 
	by Example~\ref{exmp:unstructured-J2}
	with $a_ {j} = n \tau \sigma \exp(\beta_{1} + \beta_{2} t_{j})$.
	
	For $J \geq 3$, we consider the situation of
	constant mean ability $\theta$ ($\beta_2 = 0$).
	The mean ability is the same as in 
	Examples~\ref{exmp:unstructured-noeffect}  
	and \ref{exmp:intercept-only},
	but the model is different and
	cannot be attained by reparameterization 
	because the number $p$ of parameters varies here ($p = 2$) 
	from Example~\ref{exmp:unstructured-noeffect} ($p = J \geq 3$) 
	and Example~\ref{exmp:intercept-only} ($p = 1$).
	In particular, in the associated fixed effect model
	(cf.\ Remark~\ref{rem:quasi-information-fixed}),
	the quasi-information matrix $\mathbf{M}(\xi)$ 
	is equal to the common information matrix 
	in the ordinary linear 	regression model 
	up to the constant factor $\mu = \theta \sigma$.
	For that model, it is well-known
	that the $D$-optimal design is
	the uniform two-point design $\bar{\xi}_{1, J}$ 
	which assigns equal weights $w_{1} = w_{J} = 1/2$ 
	to the first and to the last time point 
	$t_1$ and $t_J$ only.
	Hence, the design $\bar{\xi}_{1, J}$
	is also locally $D_{\mathrm{Q}}$-optimal
	for the present 
	fixed effect straight line log mean ability model.
	
	In the case of a pure permanent test person effect 
	($\rho = 1$), 
	the $D$-criterion~\eqref{eq:d-criterion-intercept}
	becomes
	$ \Phi_{D, \mathrm{Q}}(\xi)
		= \log(\det(\mathbf{M}(\xi))) - \log(1 + n \tau \mu)$,
	where $\mathbf{M}(\xi)$ is the 
	information matrix in the fixed effect model.
	Thus, also in the case $\rho = 1$,
	the design $\bar{\xi}_{1, J}$ is locally $D_{\mathrm{Q}}$-optimal
	for any combination of the other parameters
	as long as $\beta_2 = 0$.
	
	When there are random time effects 
	($\rho < 1$), it can be checked 
	by means of the equivalence theorem 
	(Theorem~\ref{thm:equivalence}~b)
	under which parameter combinations 
	the design $\bar{\xi}_{1, J}$ 
	remains to be optimal.
	For simplicity, we will discuss this in the
	situation of $J = 3$ 
	equally spaced time points $t_j = 0, 1, 2$.
	If the standardized mean response 
	$a_1 = n \tau \mu$ is small, 
	$a_1 \leq 2 (\sqrt{2} - 1) \approx 0.828$,
	then $\bar{\xi}_{1,3}$ 
	is locally $D_{\mathrm{Q}}$-optimal
	for all $\rho \geq 0$.
	When the value of $a_1$ is larger, 
	then the design $\bar{\xi}_{1, 3}$ 
	is locally $D_{\mathrm{Q}}$-optimal
	if (and only if) 
	$\rho \geq \rho_{\mathrm{crit}} 
		= 1 - 2 ((2 + a_1)^{1/3} - 1) / a_1$.
	The critical value $\rho_{\mathrm{crit}}$
	is plotted in
	Figure~\ref{fig:straight-line-j3-slope0-rhocrit}
	\begin{figure}
		\centering
		\includegraphics[width=0.55\textwidth]{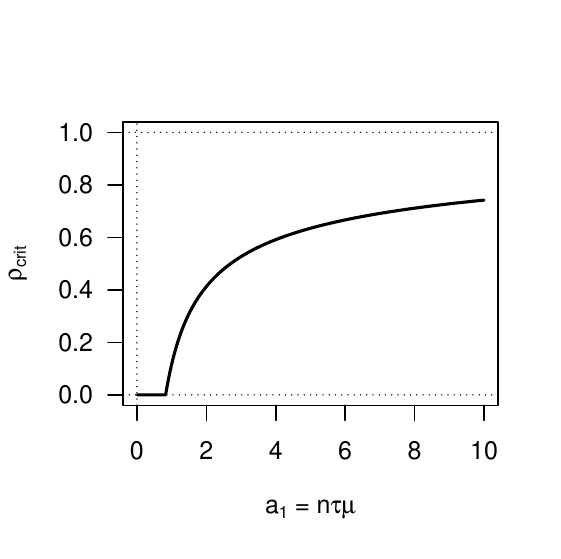}
		\caption{Critical value $\rho_{\mathrm{crit}}$
			in dependence on $a_1$
			in the straight line log mean ability model 
			($p = 2$) for $J = 3$}
		\label{fig:straight-line-j3-slope0-rhocrit}
	\end{figure}
	The vertical dotted line at $a_1 = 0$ 
	indicates the limiting case of
	the fixed effect model,
	and the horizontal dotted line at $\rho = 1$ 
	indicates the situation 
	of a pure permanent test person effect.
	For parameter combinations of $a_1$ and $\rho$
	located above (or on) the curve,
	the design $\bar{\xi}_{1,3}$ 
	is locally $D_{\mathrm{Q}}$-optimal
	while, for parameter combinations below the curve, 
	positive weight is required for the
	locally $D_{\mathrm{Q}}$-optimal design
	at all three time points
	with $w_1^* = w_3^*$ by symmetry.
	In the latter case, information will be used 
	from the interior of the time interval 
	to better estimate 
	the intercept parameter $\beta_1$.
	The critical value $\rho_{\mathrm{crit}}$ 
	of the correlation is increasing in $a_1$
	and tends to $1$ for $a_1 \to \infty$.
	Thus, when the standardized mean ability 
	$a_1$ is sufficiently large 
	and the correlation $\rho$ is not too small,
	the locally $D_{\mathrm{Q}}$-optimal design 
	assigns positive
	weight to all testing instances.
	This is, for example, the case 
	of a realistic test scenario in which 
	$a_1 = 120$ and $\rho = 0.9$ 
	(cf.\ Example~\ref{exmp:exponential-saturation}),
	where the critical value is 
	$\rho_{\mathrm{crit}} = 0.934$.
	In the special case $\rho = 0$, 
	the transition from optimal
	two-point to optimal three-point designs
	behaves like that observed by
	\citet[Figure~5]{freise2023repeated}
	in the corresponding linear model.
		
	When $\beta_2 \neq 0$,
	the situation is more complicated
	as numerical computations show
	in the case $J = 3$:
	For $\beta_2$ close to $0$,
	$\beta_2 > 0$ large, 
	or $\beta_2$ strongly negative,
	the locally $D_{\mathrm{Q}}$-optimal design 
	is a two-point design
	on $t_1$ and $t_3$ ($w_2^* = 0$),
	$t_2$ and $t_3$ ($w_1^* = 0$),
	or $t_1$ and $t_2$ ($w_3^* = 0$), 
	respectively,
	when $\rho$ is sufficiently large.
	The optimal weights 
	on the two support points 
	can be determined 
	by Example~\ref{exmp:unstructured-J2}.
	For intermediate values of $\beta_2$,
	the locally $D_{\mathrm{Q}}$-optimal design 
	assigns positive weight 
	to all three time points.
	This behavior even occurs in the situation 
	of the fixed effect model
	as could be motivated by discretization of the results in 
	\cite{rodriguez2007poisson} 
	where a continuous design region is considered.
	We will not give details here.
	\qed
\end{exmp}

\begin{exmp}[exponential saturation]
	\label{exmp:exponential-saturation}
	In the model~\eqref{eq:model-growth}
	of exponential saturation
	($\eta(t, \bm{\beta}) 
		= \beta_{1} - \beta_{2} \exp( - \beta_{3} t)$),
	we assume time points $t_j = j - 1$
	indicating an initial test ($t_1 = 0$) and
	$J - 1$ subsequent follow-up tests.
	For a reasonable test scenario,
	we consider a total number
	$n = 120$ of items per subject
	with standardized easiness $\sigma = 1$
	to be presented in $J = 3$ or $J = 7$
	testing instances.
	Concerning the dispersion of the random effects,
	we suppose $\tau = 1$ for the scale parameter 
	and a strong correlation $\rho = 0.9$.
	For the log mean ability,
	we assume nominal values $\beta_{1} = 3$
	for the saturation level, 
	$\beta_{2} = 2$
	for the range parameter,
	and $\beta_{3} = 1$
	for the speed parameter.
	Then, at time $t_1 = 0$ at baseline, 
	the initial log mean ability is
	$\eta(0, \bm{\beta}) = 1$
	and increases in time.
	Numerical values of the optimal weights
	are presented in 
	Table~\ref{tab:exponential-weights-J3-7}.
	\begin{table}
		\caption{Optimal weights
			in the exponential saturation model 
			for $J = 3$ and $J = 7$, 
			when $n = 120$, $\sigma = 1$, $\tau = 1$, $\rho = 0.9$, 
			$\beta_1 = 3$, $\beta_{2} = 2$, and $\beta_{3} = 1$}
		\label{tab:exponential-weights-J3-7}
		\centering
		\begin{tabular}{c|ccccccc}
			$J$ & $w_{1}^*$ & $w_{2}^*$ & $w_{3}^*$ & $w_{4}^*$ & $w_{5}^*$ & $w_{6}^*$ & $w_{7}^*$  
			\\
			\hline  
			$3$ &$0.510$&$0.273$&$0.217$ &&&&
			\\
			$7$ &$0.410$&$0.179$&$0.097$&$0.068$&$0.074$&$0.083$&$0.088$
		\end{tabular}
	\end{table}
	For $J = 3$, the optimal weights 
	were obtained 
	by an exhaustive search
	on a sufficiently dense grid.
	For $J = 7$, we used an augmented 
	Lagrange multiplier method 
	with an interior sequential quadratic programming algorithm 
	\citep{ye1987interior} 
	implemented in the R \citep{R2021Austria}
	package Rsolnp \citep{ghalanos2015rsolnp}
	to compute the optimal weights.
	Table~\ref{tab:exponential-weights-J3-7}
	shows that
	under the given nominal values
	a substantially larger number of items
	is to be presented at the initial time at baseline  
	($t_1 = 0$)
	than at the subsequent retesting instances
	for both $J = 3$ and $J = 7$.
	In particular, in the saturated case ($J = p = 3$),
	the optimal weight is decreasing over time.
	This is in accordance with the findings in
	Example~\ref{exmp:unstructured-J2}
	that higher weight is allocated to time points
	with lower log mean ability
	because the optimal weights depend only on
	the mean response $\mu_{1}$ to $\mu_{3}$
	and not on the model 
	equation~\eqref{eq:model-growth}.
	For $J = 7$, 
	when there are more time points than parameters,
	the monotonicity behavior does no longer hold
	because the model structure 
	may require a higher weight 
	at the last testing instance
	than at the directly preceding ones
	\citep[cf.][Theorem~1, for a related fixed effect model]{dette2010ode-exponential}.
		
	In order to assess the sensitivity of the locally optimal design
	with regard to variation of the parameters,
	we present optimal weights in
	Table~\ref{tab:exponential-weights-J3}
	for $J = 3$
	when either the
	saturation level $\beta_{1}$
	(Table~\ref{tab:exponential-weights-J3}~(a)),
	the range parameter $\beta_{2}$
	(Table~\ref{tab:exponential-weights-J3}~(b)),
	or the speed parameter $\beta_{3}$
	(Table~\ref{tab:exponential-weights-J3}~(c))
	deviates from its nominal value,
	respectively.
	\begin{table}
		\caption{Optimal weights
			in the exponential saturation model 
			for various values of $\beta_1$, $\beta_{2}$, and $\beta_{3}$
			when $J = 3$, 
			$n = 120$, $\sigma = 1$, $\tau = 1$, and $\rho = 0.9$}
		\label{tab:exponential-weights-J3}
		\centering
		\begin{tabular}{c|ccccccccc}
			\multicolumn{5}{l}{(a) $\beta_{2} = 2$, $\beta_{3} = 1$} &&&&&
			\\
			$\beta_{1}$ & $0$ & $1$ & $2$ & $\mathbf{3}$ & $4$ & $5$ & $6$ && 
			\\
			\hline  
			$w_{1}^*$ &$0.484$&$0.499$&$0.507$&
				$\mathbf{0.510}$&$0.511$&$0.512$&$0.512$ && 
			\\
			$w_{2}^*$ &$0.285$&$0.278$&$0.275$&
				$\mathbf{0.273}$&$0.273$&$0.272$&$0.272$ &&
			\\
			$w_{3}^*$ &$0.231$&$0.223$&$0.218$&
				$\mathbf{0.217}$&$0.216$&$0.216$&$0.216$ &&
		\\
		\multicolumn{5}{l}{\mbox{}} &&&&&
		\\
			\multicolumn{5}{l}{(b) $\beta_{1} = 3$, $\beta_{3} = 1$} &&&&&
			\\
			$\beta_{2}$ & $\to 0$ & $0.5$ & $1$ & $1.5$ & 
				$\mathbf{2}$ & $2.5$ & $3$ & $3.5$ & $4$ 
			\\
			\hline  
			$w_{1}^*$ &$0.333$&$0.376$&$0.420$&$0.465$&
				$\mathbf{0.510}$&$0.554$&$0.597$&$0.636$&$0.672$  
			\\
			$w_{2}^*$ &$0.333$&$0.321$&$0.307$&$0.291$&
				$\mathbf{0.273}$&$0.255$&$0.236$&$0.218$&$0.201$ 
			\\
			$w_{3}^*$ &$0.333$&$0.303$&$0.273$&$0.244$&
				$\mathbf{0.217}$&$0.191$&$0.167$&$0.146$&$0.127$ 
		\\
		\multicolumn{5}{l}{\mbox{}} &&&&&
		\\
			\multicolumn{5}{l}{(c) $\beta_{1} = 3$, $\beta_{2} = 2$} &&&&&
			\\
			$\beta_{3}$ & $\to 0$ & $0.5$ & $\mathbf{1}$ & $1.5$ & $2$ & $2.5$ & $3$ & $3.5$ & 
			$\to \infty$ 
			\\
			\hline  
			$w_{1}^*$&$0.333$&$0.452$&
				$\mathbf{0.510}$&$0.540$&$0.555$&$0.563$&$0.568$&$0.570$&$0.574$  
			\\
			$w_{2}^*$&$0.333$&$0.306$&
				$\mathbf{0.273}$&$0.250$&$0.235$&$0.227$&$0.221$&$0.218$&$0.213$
			\\
			$w_{3}^*$&$0.333$&$0.242$&
				$\mathbf{0.217}$&$0.210$&$0.210$&$0.210$&$0.211$&$0.212$&$0.213$
		\end{tabular}
	\end{table}
	In each subsection of  
	Table~\ref{tab:exponential-weights-J3},
	the column corresponding to the nominal value
	is highlighted in bold.
	For any	parameter combination, 
	we observe $w_1^* \geq w_2^* \geq w_3^*$
	which had to be expected
	because of the monotonicity property
	of the mean ability.
	In the limiting cases
	$\beta_2 \to 0$ in
	Table~\ref{tab:exponential-weights-J3}~(b)
	and $\beta_3 \to 0$ in
	Table~\ref{tab:exponential-weights-J3}~(c),
	the mean response $\mu_1$ to $\mu_3$
	will become equal
	which results in uniform optimal weights 
	$w_j^* = 1/3$ as in
	Example~\ref{exmp:unstructured-noeffect}.
	Similarly, 
	in the limiting case $\beta_3 \to \infty$ in
	Table~\ref{tab:exponential-weights-J3}~(c),
	the mean response $\mu_2$ will approach $\mu_3$
	such that the optimal weights $w_2^* = w_3^*$
	at the retesting instances.
	For all three parameters
	$\beta_{1}$ to $\beta_{3}$,
	the optimal weight $w_1^*$
	at the initial test increases with the parameter value
	and the optimal weight $w_2^*$
	at the first retest decreases. 
	This behavior is less pronounced for the
	saturation level $\beta_{1}$
	than for the range parameter $\beta_{2}$
	and the speed parameter $\beta_{3}$.
	Moreover, the optimal weight $w_1^*$
	at the last testing instance
	also tends to decrease in the parameter values
	with the exception of a slight reversal
	when $\beta_3$ gets large.
	
	Although the locally optimal designs 
	apparently may substantially differ,
	the efficiency remains surprisingly high
	at $98\%$, at least, 
	when the true parameter value differs 
	from the nominal value
	within the range of parameter specifications in
	Table~\ref{tab:exponential-weights-J3}.
	This behavior may be explained by the 
	stabilizing contribution of 
	$a_j = n \tau \mu_j$
	which dominates the influence of the
	weight $w_j$
	as long as the latter is not too small.
	
	For the case of $J = 7$ testing instances
	which seems to be more realistic for applications,
	numerical values of the optimal weights
	are exemplified in dependence
	on the speed parameter $beta_3$
	in Table~\ref{tab:exponential-weights-J7}. 
	\begin{table}
		\caption{Optimal weights
			in the exponential saturation model 
			for various values of $\beta_3$
			when $J = 7$, 
			$n = 120$, $\sigma = 1$, $\tau = 1$, $\rho = 0.9$, 
			$\beta_{1} = 3$, and $\beta_{2} = 2$}
		\label{tab:exponential-weights-J7}
		\centering
		\begin{tabular}{c|ccccccc}
			$\beta_{3}$ & $0.02$ & $0.05$ & $0.1$ & $0.5$ & $\mathbf{1}$ & $2$ & $3$ \\
			\hline
			$w_{1}^*$ &$0.213$&$0.231$&$0.256$&$0.364$&$\mathbf{0.410}$&$0.442$&$0.452$\\
			$w_{2}^*$ &$0.118$&$0.125$&$0.132$&$0.160$&$\mathbf{0.179}$&$0.182$&$0.175$\\
			$w_{3}^*$ &$0.120$&$0.125$&$0.130$&$0.126$&$\mathbf{0.097}$&$0.069$&$0.070$\\
			$w_{4}^*$ &$0.130$&$0.128$&$0.124$&$0.087$&$\mathbf{0.068}$&$0.074$&$0.075$\\
			$w_{5}^*$ &$0.118$&$0.109$&$0.102$&$0.072$&$\mathbf{0.074}$&$0.077$&$0.076$\\
			$w_{6}^*$ &$0.110$&$0.107$&$0.101$&$0.085$&$\mathbf{0.083}$&$0.078$&$0.076$\\
			$w_{7}^*$ &$0.191$&$0.176$&$0.156$&$0.106$&$\mathbf{0.088}$&$0.078$&$0.076$
		\end{tabular}
	\end{table}
	Similar to the case $J = 3$,
	the optimal design assigns 
	highest weight to the initial testing ($j = 1$). 
	This weight $w_1^*$ increases 
	in the speed parameter $\beta_{3}$.
	Conversely, the weight $w_7^*$
	at the last testing instance decreases.
	All other weights at the intermediate 
	testing instances are quite similar 
	for small values of the speed parameter.
	For larger values, more weight is assigned
	to the first retesting ($j = 2$)
	while all subsequent
	retesting instances ($j = 3, \ldots, 7$)
	hare about the same optimal weight0.
		
	For various other 
	model parameter combinations,
	numerical values of optimal weights 
	are provided in
	\citet[Section~5.2 and Appendix~A.11]{parsamaram2022thesis}.
	\qed
\end{exmp}

\section{Discussion}
\label{sec:discussion}
In the present work we have proposed a longitudinal 
model for count data in repeated item response testing.
For this model, we have developed 
the quasi-information matrix for estimating the
location parameters of the population mean ability
curve
and an equivalence theorem for the characterization
of $D$-optimal designs for the numbers of items
to be allocated over the testing instances.
As in the corresponding situation of linear, 
metric response \citep[see][]{freise2023repeated},
the quasi information
and, hence, the optimal design 
may vary with the scale parameter and 
the intra-class correlation of the subject
and time specific random effects.
Additionally, as common in nonlinear
and generalized linear models, the quasi information
and the optimal design depend on
the location parameters as well.
Thus, local optimality has been considered here.
The sensitivity of locally optimal designs 
with respect to parameter misspecification
has been investigated by computation
of the efficiency for modified parameter values.
However, it would be desirable
to also obtain optimal designs with respect
to robust optimality criteria like minimax or
weighted (``Bayesian'') criteria
which may be object of future research.

In view of the concavity (Lemma~\ref{lem:concave})
of the quasi-information matrix,
also other commonly used criteria may be considered.
For example, linear criteria 
aim at minimization of the mean squared error
$\mathrm{tr}(\mathrm{asCov}(\mathbf{L} \hat{\bm{\beta}}_{\mathrm{MQL}}))$
for estimating the linear parameter combination 
$\mathbf{L} \bm{\beta}$ for some matrix $\mathbf{L}$
where $\mathrm{asCov}$ denotes the asymptotic covariance matrix.
These linear criteria cover the $A$-criterion 
($\mathbf{L} = \mathbf{I}_p$), the $c$-criterion
($\mathbf{L} = \bm{\ell}^\top$ for some vector $\bm{\ell}$), 
and the integrated mean squared error criteria
$\int \mathrm{asVar}(\hat{\eta}(t)) \mathrm{d} t$
for estimating the log mean ability curve $\eta(t) = \eta(t, \bm{\beta})$
($\mathbf{L}^\top \mathbf{L} 
	=  \int  
		\mathbf{f}_{\bm{\beta}}(t) \mathbf{f}_{\bm{\beta}}(t)^\top 
		\mathrm{d} t$)
and
$\int \mathrm{asVar}(\hat{\theta}(t)) \mathrm{d} t$
for estimating the mean ability curve $\theta(t)$
($\mathbf{L}^\top \mathbf{L} 
=  \int \theta(t)^2 
\mathbf{f}_{\bm{\beta}}(t) \mathbf{f}_{\bm{\beta}}(t)^\top 
\mathrm{d} t$), respectively,
where $\mathrm{asVar}$ denotes the asymptotic variance.
When there is an intercept in the model,
the minimization of 
$\mathrm{tr}(\mathbf{L} 
	\mathbf{M}_{\mathrm{Q}}(\xi)^{-1} 
	\mathbf{L}^\top)$
is equivalent to the minimization of 
$\mathrm{tr}(\mathbf{L} \mathbf{M}(\xi)^{-1} \mathbf{L}^\top)$
by the representation~\eqref{eq:quasi-information-intercept}.
In contrast to the longitudinal linear mixed model
considered in~\cite{freise2023repeated}, 
the core matrix 
$\mathbf{M}(\xi) 
	= \mathbf{F}^\top \mathbf{U}(\xi) \mathbf{F}$ 
and, hence, the optimal design can depend 
here on the dispersion parameters.
This can easily be seen, e.\,g.,
in the case of $A$-optimality
for the straight line model 
of Example~\ref{exmp:straight-line-J3}
with $J = 3$ time points
where the optimal design varies with $n \tau$ and $\rho$.
However, for saturated models ($J = p$),
we can see that the inverse core matrix is
$\mathbf{M}(\xi)^{-1} 
	= \left(\mathbf{F}^\top 
			\mathrm{diag}(\mu_j w_j) \mathbf{F}\right)^{-1}
		+ (1 - \rho) n \tau  
			\left(\mathbf{F}^\top \mathbf{F}\right)^{-1}$
by equation~\eqref{eq:quasi-information-unstructured}.
Thus, for a linear criterion in a saturated model,
the optimal design does not depend on $n$, $\tau$, 
and $\rho$, and is the same as in the fixed effects model
described in Remark~\ref{rem:quasi-information-fixed}.

In the model description in Section~\ref{sec:model},
we initially assumed that, at each time point $t_j$, 
all test persons receive the same number $n_j$ of items.
In the context of approximate designs,
this requirement may be dropped.
We may allow different individual designs 
$\xi_i = (w_{i1}, \ldots, w_{iJ})$ for each test person.
Then, by concavity (Lemma~\ref{lem:concave}), 
the design
$\bar{\xi} = (\bar{w}_1, \ldots, \bar{w}_J)$,
with averaged weights
$\bar{w}_j = \frac{1}{N} \sum_{i=1}^{N} w_{ij}$, 
provides a larger standardized quasi-information matrix
($\mathbf{M}_{\mathrm{Q}}(\bar{\xi}) 
	\geq \frac{1}{N} \sum_{i=1}^{N} \mathbf{M}_{\mathrm{Q}}(\xi_i)$).
For any monotone design criterion, 
we can thus conclude 
as in~\citet[Theorem~1]{schmelter2007optimality},
that optimal designs can be found
in the class of ``single group designs''
which assign identical weights 
$w_{ij} = w_j$ to all test persons.
In so far, the initial assumption
of identical individual designs
is not restrictive.

Although, 
in view of a permanent test person effect,
it seems to be reasonable 
that an intercept is included in the model,
we suppose that concavity 
still holds even when there is none.
For example, 
in the most simple model 
$\theta(t) = \theta_{0} \exp(\beta t)$, $t = 0, 1$,
of a pre-post experiment
with known mean ability $\theta_0$ at time 
at baseline ($t = 0$)
which does not contain an intercept
to be estimated,
numerical calculations indicate
that the quasi-information is concave.
Actually, in that model, it can be seen 
that the optimal design assigns 
all items to the active time point ($w_2^* = 1$)
when there is no or only little contribution 
of the permanent test person effect 
($\rho \leq c$ where, typically, 
$c = 1 / (n \tau \sigma \theta_0 \exp(\beta/2))$ will be small).
This optimal design seems to be reasonable because the
value of the mean ability is known at initial time $t_1 = 0$.
However, when the  
contribution of the permanent test person effect 
is sufficiently large ($\rho > c$),
then the optimal design assigns positive weight 
($w_1^* = (\rho - c) / (\rho + \exp(-\beta/2))$)
to time $t_1 = 0$ at baseline.
In that case,
estimation of the parameter $\beta$ will be adjusted
by prediction of the random effect
based on items presented at baseline.
This behavior is in accordance with findings 
for the corresponding linear mixed model in
\citet[``ratio model'']{freise2023repeated}.
But here, in contrast to the linear mixed model,
the optimal weight $w_2^*$ at the active time point 
may become arbitrarily small 
when $\beta$ is large.

Finally, we have motivated the consideration 
of quasi-information by asymptotic properties
of the maximum quasi-likelihood estimator.
But, even if we would like to use a numerical
approximation of the classical maximum
likelihood estimator for statistical analysis,
the quasi-information may serve 
as a suitable approximation of the Fisher information 
\citep[see][Section~5.3]{mielke2012approximations}.
The inverse quasi-information further provides an upper bound
for the asymptotic covariance of the maximum
likelihood estimator and, hence,
a useful tool to obtain a predetermined accuracy.

\begin{appendices}
	\label{sec:appendix}

\section{Proofs and auxiliary results} 
\label{sec:proofs}

For reference, we first reproduce two 
standard results from matrix calculus.

\begin{lemma}[Matrix Inversion Lemma]
	\label{lem:matrix-inversion-lemma}
	\mbox{}
	\\ 
	$\left(\mathbf{A} + \mathbf{b} \mathbf{b}^\top\right)^{-1} 
		= \mathbf{A}^{-1} 
			- \frac{1}{1 + \mathbf{b}^\top \mathbf{A}^{-1} \mathbf{b}} 
				\mathbf{A}^{-1} \mathbf{b} \mathbf{b}^\top \mathbf{A}^{-1}$.
\end{lemma}

\begin{lemma}[Matrix Determinant Lemma]
	\label{lem:matrix-determinant-lemma}
	\mbox{}
	\\ 
	$\det\left(\mathbf{A} \pm \mathbf{b} \mathbf{b}^\top\right)^{-1} 
		= \det(\mathbf{A})\left(1 \pm \mathbf{b}^\top \mathbf{A}^{-1} \mathbf{b}\right)$
\end{lemma}

To establish the 
representation~\eqref{eq:quasi-information-representation}
of the quasi-information in Theorem~\ref{thm:quasi-information},
we first derive the  inverse 
of the individual variance covariance matrix $\mathbf{V}$
specified in equation~\eqref{eq:covariance-y_i-varying-easiness}.

\begin{lemma}
	\label{lem:covariance-y_i-inverse}
	Let	
	$\mathbf{C}_j = \mathrm{diag}(\mu_{jk}) 
		+ (1 - \rho) \tau \bm{\mu}_{j} \bm{\mu}_{j}^\top$,
	$\mathbf{a}_j = \frac{1}{1 + (1 - \rho) \tau \mu_{j\Cdot}} \mathbf{1}_{n_j}$,
	and $\mathbf{a} = (\mathbf{a}_1^\top, \ldots , \mathbf{a}_J^\top)^\top$.
	Then 
	\begin{equation}
		\label{eq:covariance-y_i-inverse}
		\mathbf{V}^{-1} 
		=
		\mathrm{diag}\left(\mathbf{C}_j^{-1}\right)
			- \frac{\rho \tau}{1 + \rho \tau \sum_{j = 1}^{J} u_j} 
				\mathbf{a} \mathbf{a}^\top . 
	\end{equation}
\end{lemma}

\begin{proof}
	Note that
	$\mathbf{V} 
		= \mathrm{diag}(\mathbf{C}_j) + \rho \tau \bm{\mu} \bm{\mu}^{\top}$.
	By the matrix inversion lemma
	(Lemma~\ref{lem:matrix-inversion-lemma}),
	\[
		\mathbf{C}_j^{-1} 
		= \mathrm{diag}(1 / \mu_{jk}) 
			- \frac{(1 - \rho) \tau}{1 + (1 - \rho) \tau \mu_{j \Cdot}}
				\mathbf{1}_{n_j} \mathbf{1}_{n_j}^{\top}  
	\]
	and
	\begin{eqnarray*}
		\mathbf{V}^{-1} 
		& = &		
		\mathrm{diag}\left(\mathbf{C}_j^{-1}\right)
		\\ 
		& &	
		\mbox{} - \frac{\rho \tau}{1 + \rho \tau \sum_{j = 1}^{J} 
				\bm{\mu}_{j}^\top \mathbf{C}_j^{-1} \bm{\mu}_{j}} 
			\mathrm{diag}\left(\mathbf{C}_j^{-1}\right) 
			\bm{\mu} \bm{\mu}^\top 
			\mathrm{diag}\left(\mathbf{C}_j^{-1}\right) \, . 
	\end{eqnarray*}
	Further, 
	$\mathbf{a}_{j} = \mathbf{C}_j^{-1} \bm{\mu}_{j}$
	and
	$u_j = \bm{\mu}_{j}^\top \mathbf{C}_j^{-1} \bm{\mu}_{j}$,
	which completes the proof.
\end{proof}

\begin{proof}[Proof of Theorem~\ref{thm:quasi-information}]
	Note that
	$\mathbf{D}_j^\top \mathbf{C}_j^{-1} 
		= \frac{1}{1 + (1 - \rho) \tau \mu_{j \Cdot}} 
			\mathbf{f}_{\bm{\beta}}(t_j) \mathbf{1}_{n_j}^\top$
	and
	$\mathbf{D}_j^\top \mathbf{C}_j^{-1} \mathbf{D}_j
		= u_j \mathbf{f}_{\bm{\beta}}(t_j) \mathbf{f}_{\bm{\beta}}(t_j)^\top$.
	Hence,
	\[
		\mathbf{D}^\top \mathrm{diag}\left(\mathbf{C}_j^{-1}\right) \mathbf{D}
			= \mathbf{F}^\top \mathbf{U} \mathbf{F} \, .
	\]
	Further
	$\mathbf{D}_j^\top \mathbf{C}_j^{-1} \bm{\mu}_{j} 
		= u_j \mathbf{f}_{\bm{\beta}}(t_j)$
	and, thus,
	$\mathbf{D}^\top \mathbf{a}  
		= \mathbf{F}^\top \mathbf{u}$.
	As
	\[
		\mathbf{D}^{\top} \mathbf{V}^{-1} \mathbf{D}
		=
		\mathbf{D}^\top \mathrm{diag}\left(\mathbf{C}_j^{-1}\right) \mathbf{D}
			- \frac{\rho \tau}{1 + \rho \tau \sum_{j = 1}^{J} u_j} 
				\mathbf{D}^\top \mathbf{a} \mathbf{a}^\top \mathbf{D}  
	\]
	by Lemma~\ref{lem:covariance-y_i-inverse},
	the result follows.
\end{proof}

Next, we establish that the quasi-information 
is increasing in the easiness of the items
in the sense of Loewner ordering,
where the relation 
$\mathbf{A} \geq \mathbf{B}$ 
for two matrices $\mathbf{A}$ and $\mathbf{B}$
means that
$\mathbf{A} - \mathbf{B}$ is nonnegative definite.

\begin{proof}[Proof of Lemma~\ref{lem:max-easiness}]
	Let $(\sigma_{jk})_{jk}$ and $(\tilde{\sigma}_{jk})_{jk}$
	the easiness of two sequences of items
	such that $\tilde{\sigma}_{jk} \geq \sigma_{jk}$,
	for all $k = 1, \ldots , n_{j}$, $j = 1, \ldots , J$.
	Denote by $\mu_{jk}$, $\mathbf{D}$, $\mathbf{V}$, 
	and $\mathbf{M}_{\mathrm{Q}}$, 
	the means, gradient, covariance matrix,
	and quasi-information
	related to $(\sigma_{jk})_{jk}$,	
	and by $\tilde{\mu}_{jk}$,
	$\tilde{\mathbf{D}}$, $\tilde{\mathbf{V}}$,
	and $\tilde{\mathbf{M}}_{\mathrm{Q}}$
	the corresponding expressions
	for $(\tilde{\sigma}_{jk})_{jk}$.
	Further, let $s_{jk} = \tilde{\sigma}_{jk} / \sigma_{jk}$ and
	$\mathbf{S} = \mathrm{diag}(s_{jk})$.
	Then $\tilde{\mu}_{jk} = s_{jk} \mu_{jk}$ and
	$\tilde{\mathbf{D}} = \mathbf{S} \mathbf{D}$ 
	by the representation~\eqref{eq:matrix-d-varying-easiness}. 
	By~\eqref{eq:covariance-y_i-varying-easiness},
	\[
		\mathbf{S} \mathbf{V} \mathbf{S} - \tilde{\mathbf{V}} 
			= \mathrm{diag}\left(s_{jk}^2 \mu_{jk} - \tilde{\mu}_{jk}\right)
			= \mathrm{diag}\left((s_{jk} - 1) \tilde{\mu}_{jk}\right)
			\geq \mathbf{0} \, .
	\]
	as $s_{jk} \geq 1$. 
	Hence, 
	$\tilde{\mathbf{V}} \leq \mathbf{S} \mathbf{V} \mathbf{S}$, 
	$\tilde{\mathbf{V}}^{-1} \geq \mathbf{S}^{-1} \mathbf{V}^{-1} \mathbf{S}^{-1}$,
	and
	\[
		\tilde{\mathbf{M}}_{\mathrm{Q}} 
			= \tilde{\mathbf{D}}^{\top} \tilde{\mathbf{V}}^{-1} \tilde{\mathbf{D}}
			\geq \mathbf{D}^{\top} \mathbf{V}^{-1} \mathbf{D}
			= \mathbf{M}_{\mathrm{Q}}
	\]
	which completes the proof.
\end{proof}

\begin{proof}[Proof of Theorem~\ref{thm:quasi-information-intercept}]
	We start with the right hand side of equation~\eqref{eq:quasi-information-intercept}.
	By the matrix inversion lemma (Lemma~\ref{lem:matrix-inversion-lemma}),
	\begin{equation}
		\label{eq:quasi-information-intercept-inverse}
		\left(\mathbf{M}(\xi)^{-1} + \rho n \tau \mathbf{c} \mathbf{c}^\top\right)^{-1}
	=
	\mathbf{M}(\xi)
		- \frac{\rho n \tau}{1 + \rho n \tau \mathbf{c}^\top \mathbf{M}(\xi) \mathbf{c}}
			\mathbf{M}(\xi) \mathbf{c} \mathbf{c}^\top \mathbf{M}(\xi) \, .
	\end{equation}
	Because 
	$\mathbf{M}(\xi) = \mathbf{F}^\top \mathbf{U}(\xi) \mathbf{F}$
	 and 
	 $\mathbf{F} \mathbf{c} = \mathbf{1}_J$,
	 we get
	 $\mathbf{M}(\xi) \mathbf{c} = \mathbf{F}^\top \mathbf{U}(\xi) \mathbf{1}_J$
	 and
	 $\mathbf{c}^\top \mathbf{M}(\xi) \mathbf{c} = \mathbf{1}_J^\top \mathbf{U}(\xi) \mathbf{1}_J$
	 such that the right hand side of
	 equation~\eqref{eq:quasi-information-intercept-inverse}
	 is equal to $\mathbf{M}_{\mathrm{Q}}(\xi)$.
\end{proof}

\begin{proof}[Proof of Corollary~\ref{cor:d-criterion-intercept}]
	The form~\eqref{eq:d-criterion-intercept} 
	of the $D$-criterion follows directly from
	the representation~\eqref{eq:quasi-information-intercept}
	of the quasi-information in
	Theorem~\ref{thm:quasi-information-intercept}
	by the matrix determinant lemma (Lemma~\ref{lem:matrix-determinant-lemma}).
\end{proof}

Before we prove the equivalence theorem
(Theorem~\ref{thm:equivalence}),
we first have to establish some relevant 
properties of the $D$-criterion.

\begin{lemma}
	\label{lem:concave}
	If {\em (a)}~$\rho = 0$, or {\em (b)}~$\mathbf{F} \mathbf{c} = \mathbf{1}_J$,
	then the quasi-information matrix 
	$\mathbf{M}_{\mathrm{Q}}(\xi)$ is concave in $\xi$, 
\end{lemma}

As a consequence, the $D$-criterion $\Phi_{D, \mathrm{Q}}$ is concave.

\begin{proof}[Proof of Lemma \ref{lem:concave}]
	First, we observe that $u_j$ is a concave function in the weight $w$, 
	$j = 1, \ldots, J$.
	Then, $\mathbf{U}$ is concave in the design $\xi$,
	i.\,e.\ 
	$\mathbf{U}((1 - \epsilon) \xi_1 + \epsilon \xi_2) 
		\geq (1 - \epsilon) \mathbf{U}(\xi_1) + \epsilon \mathbf{U}(\xi_2)$ 
	for any design $\xi_1$, $\xi_2$ and any coefficient $\epsilon$, 
	$0 < \epsilon < 1$.
	It follows that also
	$\mathbf{M}(\xi) = \mathbf{F}^{\top} \mathbf{U}(\xi) \mathbf{F}$
	is concave which proves the assertion in the case (a)~$\rho = 0$
	by Remark~\ref{rem:quasi-information-rho0}.
	
	For the case (b)~$\mathbf{F} \mathbf{c} = \mathbf{1}_J$,
	we make use of the 
	representation~\eqref{eq:quasi-information-intercept}
	of the quasi-information matrix. 
	In the case that both
	$\mathbf{M}(\xi_1)$ and $\mathbf{M}(\xi_2)$
	are nonsingular, we get
	\begin{eqnarray*}
		\mathbf{M}_{\mathrm{Q}}((1 - \epsilon) \xi_1 + \epsilon \xi_2) 
		& \geq &
		\left(\left((1 - \epsilon) \mathbf{M}(\xi_1) + \epsilon \mathbf{M}(\xi_2)\right)^{-1}
							+ \rho n \tau \mathbf{c} \mathbf{c}^{\top}\right)^{-1} ,
		\\
		& \geq &
		(1 - \epsilon) \mathbf{M}_{\mathrm{Q}}(\xi_1) + \epsilon \mathbf{M}_{\mathrm{Q}}(\xi_2)
	\end{eqnarray*}
	where the first inequality follows from
	the concavity of $\mathbf{M}(\xi)$, 
	and the second inequality follows 
	from the proof of Lemma~2
	in~\cite{schmelter2007optimality}.
	By continuity of the quasi-information matrix
	as a function of the design,
	the concavity property can be extended to
	designs with rank deficient information matrices 
	via regularization.
\end{proof}

Next, we state the directional derivative  
of the $D$-criterion.

\begin{lemma}
	\label{lem:directional-derivative}
	Let 
	$\Psi(\xi, \xi^{\prime}) 
		= \mathrm{tr}\left(\mathbf{M}(\xi)^{-1} \mathbf{F}^\top
				(\mathbf{W}(\xi^{\prime}) - \mathbf{W}(\xi))\mathbf{U}^{\prime}(\xi)
				\mathbf{F}\right)$.
	Then the directional derivative $\Psi_{\mathrm{Q}}(\xi, \xi^{\prime})$ 
	of the $D$-criterion 
	$\Phi_{D, \mathrm{Q}}(\xi)$ in the direction of $\xi^{\prime}$ is given by
	\\
	{\em (a)} $\Psi_{\mathrm{Q}}(\xi, \xi^{\prime}) = \Psi(\xi, \xi^{\prime})$
	for $\rho = 0$, and
	\\ 
	{\em (b)} $\Psi_{\mathrm{Q}}(\xi, \xi^{\prime}) 
		= \Psi(\xi, \xi^{\prime}) 
			- \frac{\rho n \tau \mathbf{1}_{J}^{\top} 
				(\mathbf{W}(\xi^{\prime}) - \mathbf{W}(\xi)) \mathbf{U}^{\prime}(\xi) 
				\mathbf{1}_{J}}{1 + \rho n \tau \mathbf{1}_{J}^{\top} \mathbf{U}(\xi) \mathbf{1}_{J}}$
	when $\mathbf{F} \mathbf{c} = \mathbf{1}_J$.
\end{lemma}
\begin{proof}
	The directional derivative of $\mathbf{U}(\xi)$ is
	$(\mathbf{W}(\xi^{\prime}) - \mathbf{W}(\xi))\mathbf{U}^{\prime}(\xi)$.
	Hence, $\log(\det(\mathbf{M}(\xi)))$ has directional derivative $\Psi(\xi, \xi^{\prime})$
	which proves case (a) for $\rho = 0$.
	
	For the case (b)~$\mathbf{F} \mathbf{c} = \mathbf{1}_J$,
	we make use of the 
	representation~\eqref{eq:d-criterion-intercept}
	of the $D$-criterion. 
	The directional derivative of
	$\log(1 + \rho n \tau \mathbf{1}_{J}^{\top} \mathbf{U}(\xi) \mathbf{1}_{J})$
	is
	\[
		\frac{\rho n \tau}{1 + \rho n \tau \mathbf{1}_{J}^{\top} \mathbf{U}(\xi) \mathbf{1}_{J}}
			\mathbf{1}_{J}^{\top} 
				(\mathbf{W}(\xi^{\prime}) - \mathbf{W}(\xi)) \mathbf{U}^{\prime}(\xi) 
			\mathbf{1}_{J}
	\]
	which completes the proof.
\end{proof}

\begin{proof}[Proof of Theorem~\ref{thm:equivalence}],
	By Lemma~\ref{lem:concave}, 
	the $D$-criterion is concave.
	Hence, we can make use of the general equivalence theorem
	as described e.\,g.\ in~\cite{silvey1980optimal}
	which states that 
	$\xi^{*}$ is $D_{\mathrm{Q}}$-optimal if and only if 
	$\Psi_{\mathrm{Q}}(\xi^*, \xi^{\prime}) \leq 0$
	for all $\xi^{\prime}$.
	
	Further, by
	Lemma~\ref{lem:directional-derivative},
	the directional derivative $\Psi_{\mathrm{Q}}(\xi, \xi^{\prime})$
	of the $D$-criterion is linear in $\xi^{\prime}$.
	Thus, we can confine to one-point designs $\xi_j$
	which assign all items to a single time point $t_j$ ($w_j = 1)$.
	As a consequence,
	$\xi^{*}$ is $D_{\mathrm{Q}}$-optimal if and only if 
	$\Psi_{\mathrm{Q}}(\xi^*, \xi_j) \leq 0$
	for all $j$.
	
	By rearranging terms, we can see that
	the directional derivative $\Psi_{\mathrm{Q}}(\xi^*, \xi_j)$
	is the difference of the left hand side 
	minus the right hand side 
	of equation~\eqref{eq:equivalence-rho0}
	in the case (a)~$\rho = 0$
	and of equation~\eqref{eq:equivalence-intercept}
	in the case (b)~$\mathbf{F} \mathbf{c} = \mathbf{1}_J$,
	respectively,
	which completes the proof.
\end{proof}

\end{appendices}
\vspace{5mm}

\noindent
\textbf{Acknowledgment.}
This work was partly supported by grant HO\,1286/6 -- SCHW\,531/15 of the 
Deutsche Forschungsgemeinschaft (DFG, German Research Foundation).
The first author recognizes the support by a PhD scholarship of the state Saxony-Anhalt (Landesgraduiertenf\"orderung).

\bibliographystyle{apalike}
\bibliography{od-reptest-count-phs}

\end{document}